\documentclass[12pt]{spieman}  
\usepackage{amsmath,amsfonts,amssymb,,mathabx}
\usepackage{graphicx}
\usepackage{setspace}
\usepackage{tocloft}
\usepackage{physics}
\usepackage{hyperref}
\usepackage{pagecolor}
\usepackage{xcolor} 
\colorlet{revA}{black}
\usepackage{lineno}

\newcommand{\lamd}{$\lambda/D$}

\title{The Analytical Performance Model and Error Budget\\ for the Roman Coronagraph Instrument}

\author[a,*]{Bijan Nemati}
\author[b]{John Krist}
\author[b]{Ilya Poberezhskiy}
\author[b]{Brian Kern}
\affil[a]{Tellus1 Scientific, LLC, 8000 Madison Blvd., Ste. D102-265, Madison, AL 35758}
\affil[b]{Jet Propulsion Laboratory, California Institute of Technology, Pasadena, CA 91109}

\cftpagenumbersoff{figure}
\cftpagenumbersoff{table} 
\begin{document} 
\pagecolor{green!0}
\maketitle

\begin{abstract}
The Nancy Grace Roman Space Telescope (``Roman''), under development by NASA, will investigate possible causes for the phenomenon of dark energy and detect and characterize extra-solar planets. The 2.4~m space telescope has two main instruments: a wide-field, infra-red imager and a coronagraph. The coronagraph instrument (CGI) is a technology demonstrator designed to help bridge the gap between the current state-of-the-art space and ground instruments and future high-contrast space coronagraphs that will be capable of detecting and characterizing Earth-like planets in the habitable zones of other stars. Using adaptive optics, including two high-density deformable mirrors and low- and high-order wavefront sensing and control, CGI is designed to suppress the star light by up to 9 orders of magnitude, potentially enabling the direct detection and characterization of Jupiter-class exoplanets. Contrast is the measure of starlight suppression, and high contrast is the chief virtue of a coronagraph. But it is not the only important characteristic: contrast must be balanced against acceptance of planet light. The remaining unsuppressed starlight must also have a stable morphology to allow further estimation and subtraction. To achieve all these goals in the presence of the disturbance and radiation environment of space, the coronagraph must be designed and fabricated as a highly optimized system. The CGI error budget is the top level tool used to guide the optimization, enabling trades of various competing errors. The error budget is based on an analytical model which enables rapid calculation and tracking of performance for the numerous and diverse questions that arise in the system engineering process. In this paper we outline the coronagraph system engineering approach and the error budget. We then describe in detail the analytical model for direct imaging and spectroscopy and show how it connects to the error budget. We  introduce a number of useful ancillary metrics which provide insight into the capabilities of the instrument. Since models always need to be validated, we describe the validation approach for the CGI analytical model. 

\end{abstract}


{\noindent \footnotesize\textbf{*}Bijan Nemati, \linkable{bijan.nemati@tellus1.com} }

\clearpage 
\setcounter{tocdepth}{2}
\tableofcontents
\clearpage 

\begin{spacing}{1}   

\section{Introduction}
\label{sect:intro}  
It is now over a quarter century since the first detection of an exoplanet: a Jupiter-mass companion to a solar type star\cite{Mayor1995}. In the intervening years, there has been rapid growth in the development of techniques for detection of exoplanets. More recently, direct imaging has become possible with the advent of a new generation of coronagraphs designed to reject the light from the host star and enable imaging and spectroscopy of companions. Planets can be directly detected using either reflected sunlight peaking in the visible band for sun-like stars, or the planet's own black-body radiation which peaks in the infrared. 

\textcolor{revA}{
For ground based instruments targeting self-luminous young hot/warm Jovian planets, there are specific advantages to working in the infrared part of the spectrum relative to the visible wavelengths. In the near IR, these planets have relatively higher planet-to-star flux ratios, depending on the spectral type of the host star, and the extreme adaptic optics (Ex-AO) instruments are also more effective than in the visible range.
Imaging the mature, cooler Jovian planets, on the other hand, is more effective in the bluer, visible part of the spectrum in reflected light. Moreover, the visible part of the spectrum offers improved characterization of even the young Jovians.\cite{Lacy2020ProspectsWavelengths}}
But imaging Jupiter-class or smaller planets around Sun-like stars, at separations resembling our solar system, will be very challenging for ground based instruments due to overwhelming atmospheric disturbances. For a space telescope, the atmospheric constraints are removed. The visible band provides access to its own set of important spectral absorption signatures, particularly valuable for characterization of exo-Earths by the next generation instruments. Operating in the visible also allows a better selection of high-performance detectors and also removes the necessity of cryogenics. 
\textcolor{revA}{Finally, for any given telescope, a shorter wavelength allows for a smaller “inner working angle” (IWA), which is the minimum angle where starlight suppression is effective.  }

The Nancy Grace Roman Space Telescope (``Roman'') is a flagship NASA astrophysics mission which contains two instruments: a wide-field imager (WFI) and a coronagraph instrument (CGI). The WFI is designed to take advantage of the telescope's wide field of view and low-temperature capability to measure the three-dimensional distribution of millions of galaxies (using redshift and position angles) as well as a large number of supernovae. These measurements will be used to constrain the answers to a number of important questions, including whether the cosmic acceleration arises from a breakdown of general relativity or from a new form of energy. The WFI will use gravitational microlensing to conduct a statistical survey of exoplanets over an important and complementary part of the exoplanet phase space (notably semi-major axis) compared with Kepler's transit based survey. The second Roman instrument, the CGI, is a visible-band coronagraph technology demonstrator that will narrow the gap of over four orders of magnitude between the sensitivity needed to detect an exo-Earth and the current state of the art. \textcolor{revA}{
CGI will do this by improving starlight suppression overall, and particularly close to the star. 
\cite{Poberezhskiy2022RomanStatus, Poberezhskiy2021RomanConcept, Mennesson2021TheMissions, Mennesson2020}}

For a coronagraph to effectively reshape and control unwanted starlight diffraction, the ideal pupil would be an unobscured circular aperture because of its symmetry and mathematical simplicity. Thus, the integration of a coronagraph to a wide field telescope like Roman, with its large secondary mirror and struts, posed a number of challenges to the optical design. A trade study between the best known architectures in 2014 led to the selection two concurrent architectures, a Hybrid Lyot Coronagraph (HLC) and a Shaped Pupil Coronagraph (SPC).\cite{Krist2015b} Each of these approaches has its own strengths and both are part of the CGI baseline design. With the HLC, the region of highest starlight suppression, the so-called ``dark hole'' (DH) is circularly symmetric and reaches close to the star. The SPC, on the other hand, has lower chromatic error and pointing sensitivity. In all, CGI will use three coronagraphs: HLC for direct narrow-field imaging, SPC (used in conjunction with the prism-slit spectrograph) for narrow-field spectroscopy, and a different SPC design with a circularly symmetric dark hole for wide-field imaging.\cite{Riggs2021FlightInstrument}

In the context of exoplanet imaging, sensitivity is best described in terms of a planet's flux ratio, defined as the ratio of the flux received from the planet over the flux received from its host star. A planet's reflected-light flux ratio can be estimated if we have a model of the albedo and phase function:
\begin{equation}
\label{eq:fluxRatio}
\xi = A_g\ \phi(\alpha) \left(\frac{R_p}{a}\right)^2 \, ,
\end{equation}
where $A_g$ is the geometric albedo, $\phi(\alpha)$ is the phase function, $R_p$ is the planet radius, and $a$ is the distance from the planet to the star. 
\textcolor{revA}{The flux ratio is not only the important parameter in determining the signal strength expected from the planet, it is also the scientifically interesting quantity, particularly for planets detected in reflected light. As Eq.~\ref{eq:fluxRatio} suggests, key planet characteristics such as its radius, semi-major axis and albedo all affect the flux ratio.}
Using the Lambertian sphere approximation,\cite{Traub2010b} the phase function is given by: 
\begin{equation}
\label{eq:LamberPhaseFun}
\phi(\alpha) = 
\frac{1}{\pi}\left[\ \sin(\alpha)+
(\pi-\alpha)\cos(\alpha)\ \right]   ,
\end{equation}
where $\alpha$ is the phase angle. At quadrature phase (i.e.,``half-moon"), $\phi(\pi/2)=1/\pi$. 
\textcolor{revA}{A more optimal phase angle for reflected light imaging is $\sim65^\circ$, where $\phi(\alpha)\simeq0.56$. At this phase angle, an exo-Jupiter observed in the visible band with  a geometric albedo of $0.54$\cite{Mallama2017a} has a flux ratio of $2.6\times10^{-9}$ or 2.6 part per billion (ppb). }

The CGI threshold (minimum) requirement is that is must detect a planet with flux ratio of $10^{-7}$ (100 ppb) with a signal to noise ratio (SNR) of at least 5. The CGI aspirational performance goal is up to two orders of magnitude more ambitious, down to the detection of a 5 ppb flux ratio planet like 47 UMa c, \textcolor{revA}{which is a potentially detectable, known Jovian exoplanet, with an orbit similar to Jupiter, around a Sun-like star.\cite{Fischer2002AMajoris}} 
Additionally, the CGI is equipped with a prism-slit spectrograph for exoplanet spectroscopy with a resolving power of $R=\lambda/\delta \lambda = 50$. The CGI will be a particularly powerful instrument for disk detection.

The CGI has been designed to support a number of coronagraphic mask configurations. The baseline operating modes are listed in Table~\ref{tab:CGIobsModes}. A more detailed description of CGI modes is provided elsewhere.\cite{Riggs2021FlightInstrument}.

\begin{table}[ht]
\textcolor{revA}{
\caption{The primary CGI observational modes. Abbreviations used are: IMG for direct imaging and photometry, SPEC for spectroscopy, NFOV for narrow field of view, and WFOV for wide field of view. $\lambda_c$ is the band central wavelength, WA is working angle, and Az is short for Azimuthal. }
\label{tab:CGIobsModes}
\begin{center}
\resizebox{\textwidth}{!}{
\begin{tabular}{|l||l|c|c|c|c|c|l|}  
\hline
\rule[-1ex]{0pt}{3.5ex}  CGI Mode & Coronagraph & $\lambda_c$, nm & BW & WA, $\lambda/D$& FOV, mas  & Az FOV& Use case  \\
\hline \hline 
\rule[-1ex]{0pt}{3.5ex} NFOV Band 1  & HLC & 575  & 10\% & 3-9  & 151-487 &  360$^\circ$ & IMG    \\
\hline
\rule[-1ex]{0pt}{3.5ex} NFOV Band 3 & SPC Bowtie & 730  & 15\% & 3-9 & 173-524 &  130$^\circ$ & SPEC, R=50    \\
\hline
\rule[-1ex]{0pt}{3.5ex} WFOV Band 4  & SPC Wide & 824  & 10\% & 6-18 & 425-1447 & 360$^\circ$ & IMG    \\
\hline
\end{tabular}}
\end{center}  }
\end{table}

\textcolor{revA}{
An important attribute of any coronagraph is its \textit{working angle} range. Working angle (WA) is measured in units of collecting aperture diffraction limit. For example, for Roman, the collecting aperture diameter $D$ is 2.36~m, so in Band 1, where the central wavelength $\lambda$ is 575~nm, the diffraction limit \lamd\ works out to be about 50~mas. The table shows that for Band 1 the IWA is 3 \lamd, or   151~mas. With this small an IWA, an exo-Jupiter would still be within the dark hole even out to $\sim$20 pc distance. This is important because future aspirational targets for the next generation space telescopes will be exo-Earths (e.g. with 100 mas separation at 10 pc). A small IWA for future coronagraphs will be critical for realizing this goal, and CGI is an important step towards demonstrating this capability.  }

\subsection{Contrast and Core Throughput Defined}

Any technical discussion of high contrast imaging must include the fundamental concepts of contrast and throughput. These are natural to the goal of maximizing SNR, since contrast is concerned with the  {\em suppression of the background} (starlight) while the off-axis throughput is concerned with {\em acceptance of the signal} (planet light). Historically, ``contrast'' has been defined as the measure of the intrusion of starlight into the planet signal area, with the unfortunate consequence that ``high  contrast'' refers to the condition where the contrast is in fact numerically small. Nevertheless, to avoid confusion we will follow the customary sense of this metric.

\begin{figure}[ht]
\centering
\includegraphics[width=0.9\columnwidth]{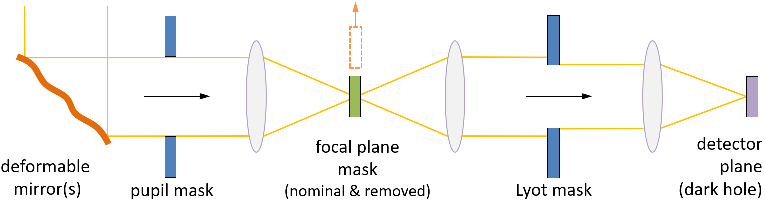}
\caption{ Basic elements of a coronagraph. } 
\label{fig:basicCG}
\end{figure} 

We begin with contrast. Consider a case where the telescope is pointed at a star, and there are no planets: we are only concerned at this point with how much starlight is scattered into an off-axis region of interest. The light incident on the primary is compressed by the telescope and sent into the coronagraph optical system (Fig.~\ref{fig:basicCG}). A small fraction of the starlight diffracts to a given region of interest in the final image plane. Noting that there is a direct correspondence between the image plane and points on the sky, we use the same coordinates for the corresponding points in the definitions that follow. We consider, as the region of interest, an area centered at $(u,v)$ and of size $\Omega_{m}$  (the corresponding solid angle on the sky). \textcolor{revA}{$\Omega_{m}$ can be a single pixel or a cluster of contiguous pixels representing the core region of the PSF. The important thing is that the same $\Omega_m$ be used consistently in the calculations.}
Let $\tau_{\ell}(u,v)$ represent the fraction of the ``leakage'' from the star, which is located at $(0,0)$ on the sky, that ends up in the $\Omega_{m}$-sized region of interest at  $(u,v)$ on the image plane. This is in effect a throughput.  Separately, let $\tau_{pk}(u,v)$ represent the ``peak'' throughput, for the case where the star is not at $(0,0)$ but at $(u,v)$, the center of the region of interest. 
We define contrast at $(u,v)$  as:
\begin{equation}
\label{eq:contrast}
C(u,v)\equiv\frac{\tau_{\ell}(u,v)}{\tau_{pk}(u,v)} \,.
\end{equation}
Note that with this definition, the size (or even the shape) of the region of interest is common to the denominator and numerator: changing $\Omega_m$ only changes the size of the region over which the contrast is averaged.

\begin{figure}[ht]
\centering
\includegraphics[width=1.0\columnwidth]{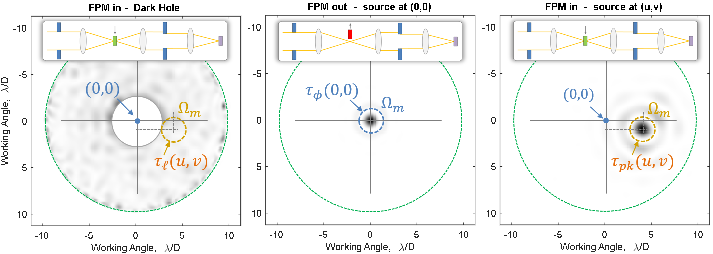}
\caption{
\textcolor{revA}{
Both Contrast $(C)$ and Normalized Intensity $({\rm NI})$ require obtaining the dark hole leakage $\tau_{\ell}$ (left), summed over a specified region of interest $\Omega_m$ centered  at $(u,v)$. For NI, this is normalized by the peak throughput $\tau_{\phi}$ (center), evaluated over the same sized  $\Omega_m$, centered at $(0,0)$ with no focal plane mask (FPM) present. For contrast, it is normalized by the peak throughput $\tau_{pk}$ (right) for an \textit{off-axis} source, centered at $(u,v)$, summed over  $\Omega_m$. }
}
\label{fig:CvsNI}
\end{figure}

In practice, evaluating contrast, whether in a model or in actual hardware, is time intensive and expensive. In Eq.~\ref{eq:contrast}, evaluating $\tau_{\ell}(u,v)$ is relatively straightforward: a point source is placed at the field point $(0,0)$ and the light everywhere in the final image plane (all $(u,v)$)  is measured or computed at once. 
\textcolor{revA}{Figure~\ref{fig:CvsNI} illustrates the components of both C and NI. }

Measuring $\tau_{pk}$ requires a  {\em separate} model propagation or testbed/instrument input wavefront tilt for each $(u,v)$ of interest.
Because of this, a more convenient, approximate measure is also used, called {\em normalized intensity} (NI). NI is defined similarly to contrast, except the denominator $\tau_{pk}(u,v)$ is replaced by $\tau_\phi(0,0)$, the throughput into the $\Omega_m$-sized region at $(0,0)$ {\em with no focal plane mask present}: 
\begin{equation}
\label{eq:NI}
{\rm NI}(u,v)\equiv\frac{\tau_{\ell}(u,v)}{\tau_{\phi}(0,0)} \,.
\end{equation}

Except for the mask removal (indicated in Fig.~\ref{fig:basicCG} and Fig.~\ref{fig:CvsNI} (middle pane) with a removed focal plane mask), $\tau_{\phi}(0,0)$ is evaluated identically to $\tau_{\ell}(u,v)$. In particular, it sums over the same $\Omega_m$-sized ROI. 

Two cautionary points are in order with NI. First, contrast is the metric of actual interest, and NI is an approximation. In the literature, some authors use the term ``contrast'' when the quantity calculated is in fact NI. This is important because of the second cautionary point: for smaller separations from the host star, NI tends to be numerically smaller (looks ``better'') than contrast, by a factor of a few. This is because $\tau_{pk}(u,v)$ drops at small separations
\textcolor{revA}{as the planet PSF get clipped by the focal plane mask,}
while  $\tau_{\phi}(0,0)$ experiences no such loss. Often the smaller separations are exactly where most of the targets are, so this can be a significant error. The remedy is to base all calculations on contrast, and account for the contrast-to-NI ratio separately via modeling. We can define this ratio as:
\begin{equation}
\label{eq:c_over_NI}
c(u,v)\equiv\frac{C(u,v)}{{\rm NI}(u,v)}=\frac{\tau_{\phi}(0,0)}{\tau_{pk}(u,v)} \,.
\end{equation}
Usually an annular average, \textcolor{revA}{$\bar{c}(r)$}, will suffice to characterize $c$.

\begin{figure}[h]
\begin{center}
\begin{tabular}{c}
\includegraphics[height=5.5cm]{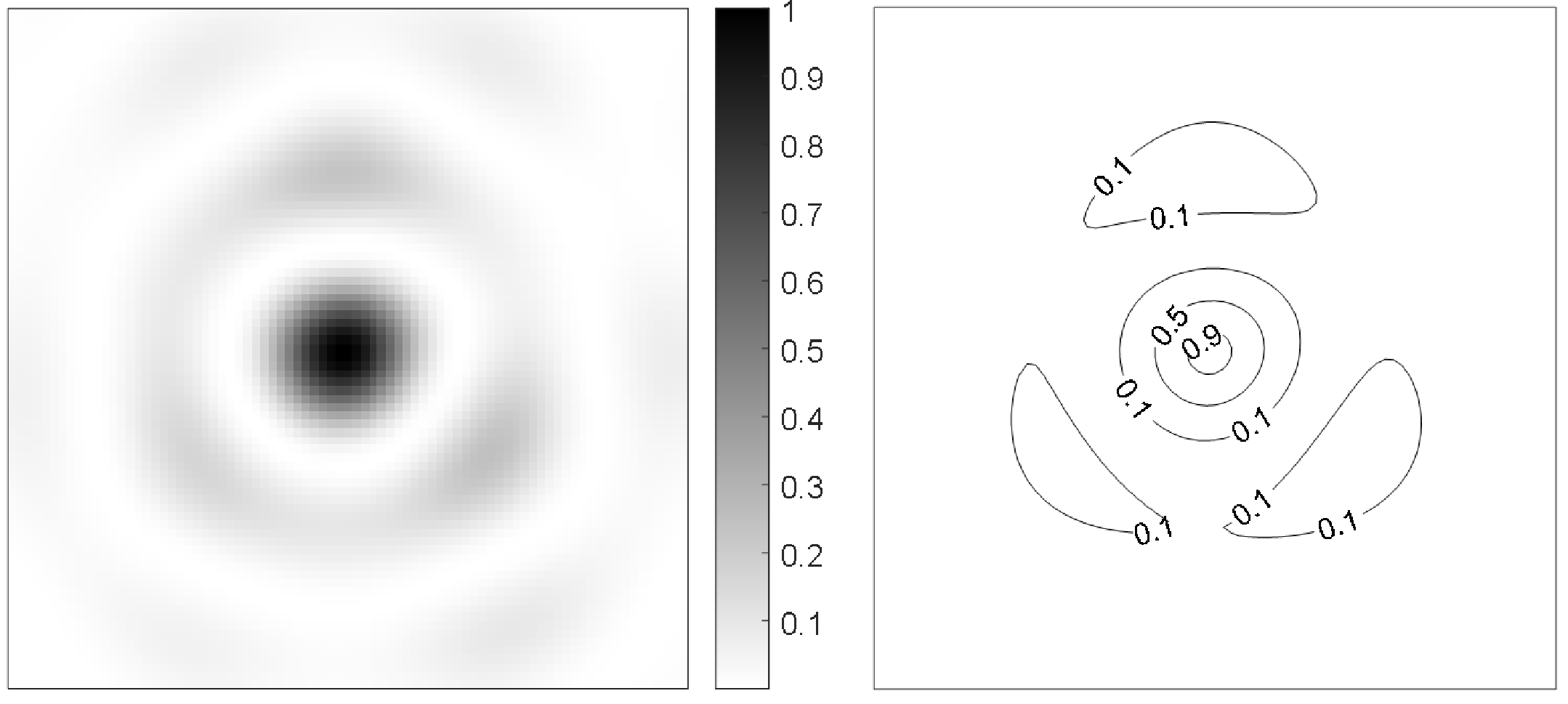}
\end{tabular}
\end{center}
\caption{ \label{fig:PSFcore}
\textcolor{revA}{
The Roman CGI PSF for the HLC coronagraph mode at 3.5 \lamd, shown in grayscale (left), and contour plot(right). The core is the area inside of the 0.5 contour.}} 
\end{figure}

The planet signal will extend over a broad region, according to the point spread function (PSF) of the optical system. In the analytical model that we will be describing, we assume a simple aperture photometry approach, where the light from the bright, or ``core'' part of the PSF will provide the signal. We assume that all the light in the detector pixels that are within the PSF core contribute both to the signal and the noise. The signal extraction in the real analysis will likely involve some form of matched filter in order to minimize noise, but most of the SNR gain is from the core region, and a properly sized aperture is not far from the optimum. 
We define the planet \textit{PSF core} as the area on the image plane circumscribed by the PSF half-max contour. An example of the HLC PSF is shown in Fig.~\ref{fig:PSFcore} with the core area also indicated.

We define the {\em core encircled energy}, $\tau_{\rm PSF}$, as the fraction of the PSF energy diffracted into the core region. For a perfect Airy disk, the core encircled energy is $\sim$49\%, but for the coronagraph, the number depends on the coronagraph design and the separation from the star.  For the HLC and a source at 4 \lamd\ working angle, the core encircled energy is $\sim$19\%. The full accounting of the diffractive effects requires another factor, which we call the {\em occulter transmission}, $\tau_{occ}$. It is defined as the fraction of the incident light on the (unobscured) collecting area that ends up in the final image plane. 
The full effect of diffraction on the throughput is summarized by the {\em core throughput}:
\begin{equation}
\label{eq:tau_core}
 \tau_{core}=\tau_{\rm PSF}\cdot\tau_{occ} \, .
\end{equation}

Core throughput is an important measure of the efficiency of a coronagraph in transmitting the planet light. The best coronagraph has high core throughput and high (numerically small) contrast.  In the hardware the detector pixel is \textcolor{revA}{where the focal plane intensity is sampled}, while in diffraction modeling the model pixel size fills that role. 
\textcolor{revA}{Because diffractive effects are band limited, light incident on the focal plane does not possess any ``structure'' finer than \lamd\ in size (projected to the pupil). Hence, it makes sense to define the \textit{sampling} $s$ of a pixel in the focal plane as:}
\begin{equation}
\label{eq:sampling}
 s=\frac{\theta_{pix}}{\lambda/D} \, .
\end{equation}
where $\theta_{pix}$ is the pixel plate scale on the sky. Thus, for example, the Nyquist sampling condition is met when $s\leq0.5$. The sampling of the modeling pixels is usually chosen to be finer than that of the detector pixels. In aperture photometry, where the region of interest is the PSF core located at the planet location, the starlight background (speckle) has the throughput $\tau_{\ell}(u,v)\ m_{cor}$ where $m_{cor}$ the number of pixels subtended by the PSF core.

\subsection{Analytical Model Description in this Paper}

Success in reaching the performance objectives requires an integrated system engineering process. The Roman CGI technology program, completed by early 2017, was successful in demonstrating the starlight suppression, wavefront sensing and control, and faint signal detection in the presence of flight like disturbances. Additionally, an extensive integrated modeling program has guided the design. The integrated model, which has been described in detail elsewhere, has been validated against the testbed results and has reached a high degree of maturity.\cite{Krist2015b} In addition to the integrated model, CGI uses a detailed analytical model to estimate the coronagraph performance in a number of ways. While the analytical model is only an approximation to the full integrated model, it in fact offers important complementary advantages. The first is that it is analytical: the physics and statistical approximations are perspicuous and can be used to build intuition in understanding how to think of the system. The second is speed: answers can be obtained in seconds instead of hours and days, over time making a significant difference in the project work flow. Finally, the analytical model is tightly integrated into the error budget, which is used to set requirements on the sub-components. 

The goal of this paper is to describe the analytical model that undergirds the CGI error budget, a work that has already informed studies of future exoplanet missions.\cite{Nemati2020b} Since every error budget needs to be based on a concept of operations, Section 2 lays out the operational concept assumed in this error budget.  Section 3 derives the photometric SNR and defines some key variables associated with the CGI operational concept. The error budget has three major branches, and these are covered in the subsequent three sections. Section 4 develops the calculations that go into the photometric branch. Section 5 covers the contrast and contrast stability branch. Section 6 describes the calibration branch. Section 7 pulls together all the foregoing material to show the top level CGI error budget. Section 8 describes the validation of the error budget analytical model against the full integrated model for the coronagraph. Section 9 provides a summary and conclusion. 

\section{CGI Concept of Operation}
\label{sect:conops}

The first step in the derivation of requirements is to define the operational concept and performance metrics within that concept. CGI will be housed in the instrument carrier section of the Roman observatory, alongside the primary science instrument, the WFI. The observatory will be orbiting in the second Lagrange point (L2), where the thermal environment will be stable.  At this distance, the Earth will not cause significant eclipse events, nor will it exert a significant radiative heat load on the flight system. Time will be allocated to WFI and CGI operations where each instrument will alternately have precedence over observatory operations and resources. Once a WFI operational slot nears completion, CGI begins preconfiguring for the upcoming observation. The precision alignment mechanisms (PAM's) are commanded to the configuration needed for the observation.\cite{Riggs2021FlightInstrument}
For example, during narrow-angle direct imaging, the focal plane alignment mechanism will bring the hybrid Lyot coronagraph (HLC) mask into position, centered on the first image plane, the Lyot stop alignment mechanism will likewise position the Lyot mask into place, and the filter selector will place the appropriate filter in the light path. Deformable mirrors (DM1 and DM2), will have been commanded to the last known optimal shape for this mode starting weeks before, while WFI was still operating. This way the transient creep of the DM's lead magnesium niobate (PMN) material will have settled.\cite{Poberezhskiy2022RomanStatus} The finer modifications of the shapes during wavefront control will be sufficiently small in amplitude to minimize creep during operation. The attitude control system (ACS) is commanded to point the telescope to a pre-defined sky position, centered on a particular star. This is usually the bright reference star that will be used to created the dark hole for this observation. The ACS brings the star within the field of view of CGI's exoplanetary (``science'') camera (EXCAM). The EXCAM is then used to measure the needed pointing correction, which is fed back to the ACS. With this correction, the star will be sufficiently close to the center of the focal plane for its light to be captured by the  low-order wavefront sensor (LOWFS). The LOWFS also uses an EMCCD like the EXCAM, which is called the LOCAM.\cite{Shi2017}  

Wavefront errors can be characterized in terms of polynomials of various spatial frequency orders, measured in cycles per aperture. One very common basis set consists of the Zernike polynomials, which are ortho-normal on a circular unobscured disk. A separate invention of the Dutch physicist Frits Zernike, for which he won the Nobel prize in 1953, is the phase contrast microscope, and this is the basis of the Zernike wavefront sensor employed in the LOWFS.\cite{Poberezhskiy2021RomanConcept} 
The LOWFS can detect tip-tilt at high rate (1000 Hz) and low order wavefront errors (Zernike modes 4 through 11 in the Noll numbering scheme) at a much lower rate (0.1 Hz). As the control loop locks the pointing, it sends updates to the ACS system at a slow rate in order to refine the ACS pointing, all the while maintaining precision pointing with the fast steering mirror (FSM) using the LOWFS error signal. Focus error is corrected via a dedicated focus control mirror while all higher order modes are corrected via the deformable mirrors. DM1 is at a pupil, so that it directly controls phase, while DM2 is about 1 meter away from the pupil, allowing for a modest level of amplitude control. 

With the low-order control established, high-order control commences using an iterative process of estimation of the electric field (approximated as a complex scalar field) and application of  a wavefront and amplitude change that would cancel this residual field. Each DM's actuators and phase sheet are subject to physical constraints, such as maximum actuator displacement ("stroke limit'') and maximum difference between adjacent actuator heights (``neighbor rule''). These are all incorporated into the solution using the method of electric field conjugation (EFC), described in detail elsewhere.~\cite{Giveon2007,Giveon2011} One important aspect of this method is the amplification of the electric field.
Amplification of the field for detection purposes can be effected by adding a known additional field via the deformable mirrors. When an additional electric field $\Delta E$ is coherently mixed with and existing field $E$, the intensity of the combined field becomes:
\begin{equation}
\label{eq:fov}
|{\rm E}+\Delta {\rm E}|^2 = |{\rm E}|^2 + |\Delta {\rm E}|^2 + 2\Re({\rm E}^* \Delta {\rm E})  
\end{equation}
That is, the addition of a large coherent field to the existing dark hole field causes heterodyne amplification of the small existing field via the cross term. EFC uses a specific series of changes ($\Delta {\rm E}$ patterns over the dark hole area) to enable the estimation of the real and imaginary parts of the complex field ($\rm E$) in the image plane.
Using the known response of the field to DM actuator movements, the DM actuators are commanded to further suppress the remaining field. In each step, the dark hole field becomes weaker, and the light level drops. The final stages of the process take the most time because of the ultra low light levels. It is for this reason that the dark hole is created using a bright star and not necessarily the target star. 

Once the dark hole has been achieved, the observatory is commanded to slew (typically by many degrees) to the target star for direct imaging or spectroscopy. The configuration of the two DM's is frozen during this maneuver. The acquisition of the target star proceeds in a manner similar to the reference star, culminating in the pointing being locked on the star using the LOWFS and the FSM. Low order errors are sensed on the target star and  corrected as before, but no higher-order corrections are attempted because the target star is typically too dim.

\textcolor{revA}{
In much of the discussion that follows, we are interested in the error due to various sources, where by ``error'' we usually mean an ensemble standard deviation. That is, if we were to conduct this same visit or measurement a number of times, with the various error sources allowed to vary according to their governing physical laws, what is the standard deviation, across the ensemble, of the signal values we would get.
}

\textcolor{revA}{
During the integration that yields the target star image, there will be signal counts (electrons) along with background counts. The background will include the remaining speckle  along with zodiacal dust background from both the local- and exo-zodi, and finally also detector noise. As integration time $t$ is increased, random noise goes down as $1/\sqrt{t}$ relative to the signal. This is because the signal grows proportionally with $t$, while random noise (measured as a standard deviation) grows only as $\sqrt{t}$. 
}

But the speckle background, relative to the signal, does not necessarily go down. The speckle background does not have a temporally random nature, as it follows the  changes in the optical configuration of the system. The configuration includes geometry, DM shapes, and any changes in the refractive properties of the transmissive optics. For a well-designed system in a thermally stable environment, these changes are ideally small, but in practice they are always present. 
Since the speckle background is not temporally random, it 
\textcolor{revA}{grows}
at the same rate as the planet image with increasing integration time. 
The speckle's spatial non-uniformity, particularly any azimuthal non-uniformity, will tend to create false positives.\cite{Shaklan2011b}

\begin{figure}[ht]
\centering
\includegraphics[height=5cm]{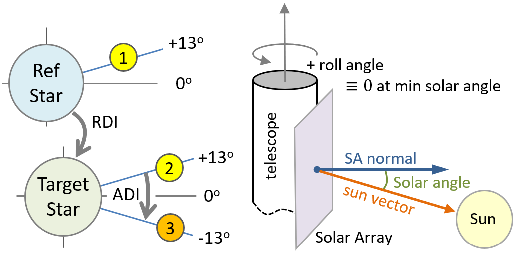}
\caption  
{ \label{fig:RDI_ADI}
In the CGI standard observing scenario, the dark hole is created in the visit to the reference star. Chops between the reference and targets star are used for RDI. The target observations are split into two roll states for ADI.}
\end{figure}

There are four major ways the noise floor can be driven lower. These are: 1) real-time monitoring of the dark hole (for example in an out-of-band channel), 2) temporal decoupling by modulating the speckle, 3) spatial smoothing, and 4) differential imaging. \textcolor{revA}{For its direct imaging technology demonstration,} Roman is planning to employ the latter two methods: a combination of reference differential imaging (RDI) and angular differential imaging (ADI), as shown in Fig.~\ref{fig:RDI_ADI}. 

RDI involves taking reference images of the speckle pattern from the reference star, typically chosen to be a bright star to minimize time consumption. The reference images are used to estimate and subtract the speckle background. ADI uses roll to average and smooth out the background prior to subtraction. But roll is constrained on Roman. Due to solar power generation constraints, the solar angle (between the solar array normal and the vector to the sun) is minimized. At zero roll angle, these two vectors lie in the same plane. When the telescope optical barrel assembly (OBA) is rolled, this angle increases. The operational requirement is that only rolls within $\pm15$ degrees are permitted. Our observing scenarios so far have kept the ADI roll to $\pm 13$ degrees to allow for more conservative performance estimates. At the HLC inner working angle of 3\lamd, the ADI roll is just barely enough to separate the footprint of the planet PSF core at the two roll positions. It is possible that the ADI roll might offer up to a $\sqrt{2}$ reduction of the background speckle, but the detailed study of what might be expected from ADI and any other post processing has been deferred to the community participation program phase of the mission. As a result, the error budget calculations in fact \textit{do not} assume ADI, and any improvements that might be provided by it, providing a level of conservatism in the performance estimates. Thus, in the following discussion we will only focus on RDI, where the speckle pattern as observed with the reference star is used as an estimate of the speckle pattern that exists while imaging the planet around the target star. Operationally, the basic steps of a minimal RDI observing scenario consist of:
\begin{enumerate}
\item observation of the reference star and generation of the dark hole,
\item slewing to, acquisition, and observation of the target star (its planets),
\item formation of a differential image.
\end{enumerate}
In the next section we describe the analytical model for differential imaging.

\section{Reference Differential Imaging SNR}

Roman CGI requirements call for a planet of a given flux ratio and separation from a star to be imaged with SNR greater than a specific value after a specified time of integration, nominally 10 hours. This amounts to a requirement on the noise in the differential image after the integration time. 
Thus, the analytical model must calculate the average SNR achievable after a given integration time, and conversely,  the average integration time needed to reach a desired SNR.
The photometric SNR is given by:
\begin{equation}
\label{eq:photometricSNR}
 {\rm SNR}=\frac{S}{\sqrt{S+\sigma_B^2} } \, .
\end{equation}
The denominator is the error in the signal measurement, coming from the shot noise of the planet itself as well as from the background. Since shot noise follows Poisson statistics, its variance equals the mean, $S$.

\subsection{The Flux Ratio Factor  and Flux Ratio Noise (FRN)}
\label{sec:Signal}

The quantity of scientific interest in direct imaging is the flux ratio. The planet signal $S$ can be written as: 
\begin{equation}
\label{eq:signalPlanetRate}
 S=r_{pl}\ t\, ,
\end{equation}
where $t$ is the integration time on target and $r_{pl}$ is the planet count rate. Planet light passes through the optical system and has a PSF when it is incident on the focal plane array (FPA). The FPA for Roman is a Teledyne e2v CCD311 electron multiplication CCD (EMCCD), which is a flight customization of the Te2v CCD201 commercial sensor.\cite{Harding2015b} 

The planet count rate is given by a product that starts with the host star flux, includes the flux ratio of the planet, the collecting area of the telescope, the attenuation of the signal as it passes through the optics, and culminates in the conversion efficiency at the detector:
\begin{equation}
\label{eq:r_pl}
 r_{pl}=F_{\lambda}\ \Delta\lambda\ \xi_{p}\ A_{col}\ \tau_{pl}\ \eta \, ,
\end{equation}
where $F_{\lambda}$ is the host star average spectral flux within the band of interest, in photons per second per area per wavelength bandwidth, and $\Delta\lambda$ is the filter wavelength bandwidth. We can alternatively think of $F_{\lambda}\ \Delta\lambda$ as shorthand notation for the integral of the flux over the band of interest.  $A_{col}$ is the effective telescope collecting area after obscurations while $\tau_{pl}$ and $\eta$ are the losses due to throughput and detector quantum efficiency, respectively. $\eta$ includes not only the photon-to-electron conversion rate but many other effects which will be discussed in Subsection~\ref{subsec:NetQuantum}. Finally, $\xi_{p}$, the objective of the measurement, the planet's flux ratio. 

The planet throughput, $\tau_{pl}$, includes the diffractive loss captured in the core throughput $\tau_{core}$ (Eq.~\ref{eq:tau_core}) along with losses from reflections and transmissions  ($\tau_{rt}$), filters ($\tau_{fil}$), and, when present, polarizers ($\tau_{pol}$): 
\begin{equation}
\label{eq:planetThroughput}
 \tau_{pl} = 
 \tau_{core} \cdot \tau_{rt} \cdot \tau_{fil} \cdot \tau_{pol} \, .
\end{equation}

Combining Eqs.~\ref{eq:signalPlanetRate} and \ref{eq:r_pl}, the conversion from the detected signal $S$ to the planet flux ratio is given by:
\begin{equation}
\label{eq:xi_p}
\xi_{p}  = \kappa \cdot S \, ,
\end{equation}
where we have defined the {\em flux ratio factor}, $\kappa$, as:
\begin{equation}
\label{eq:kappa}
\kappa \equiv \left(  F_{\lambda}\ \Delta\lambda\ A_{col}\ \tau_{pl}\ \eta\ t \right)^{-1} \, .
\end{equation}
Note that $\kappa$ depends on attributes of the observing scenario as well as the instrument. 

Since the measurement is ultimately in terms of flux ratio, it is the noise in the flux ratio that is of interest. We thus define {\em flux ratio noise} (FRN) as the error metric of the coronagraph, on which the error budget is based. 
Since the conversion is a simple proportionality factor, there is a direct correspondence (i.e. $\kappa$) between noise counts in the differential image from which the signal is extracted, and the noise that applies to the final photometric result (the measured planet flux ratio).

\subsection{Noise in the RDI image}\label{sec:NoiseInRDIImage}

\begin{figure}[ht]
\centering
\includegraphics[height=5cm]{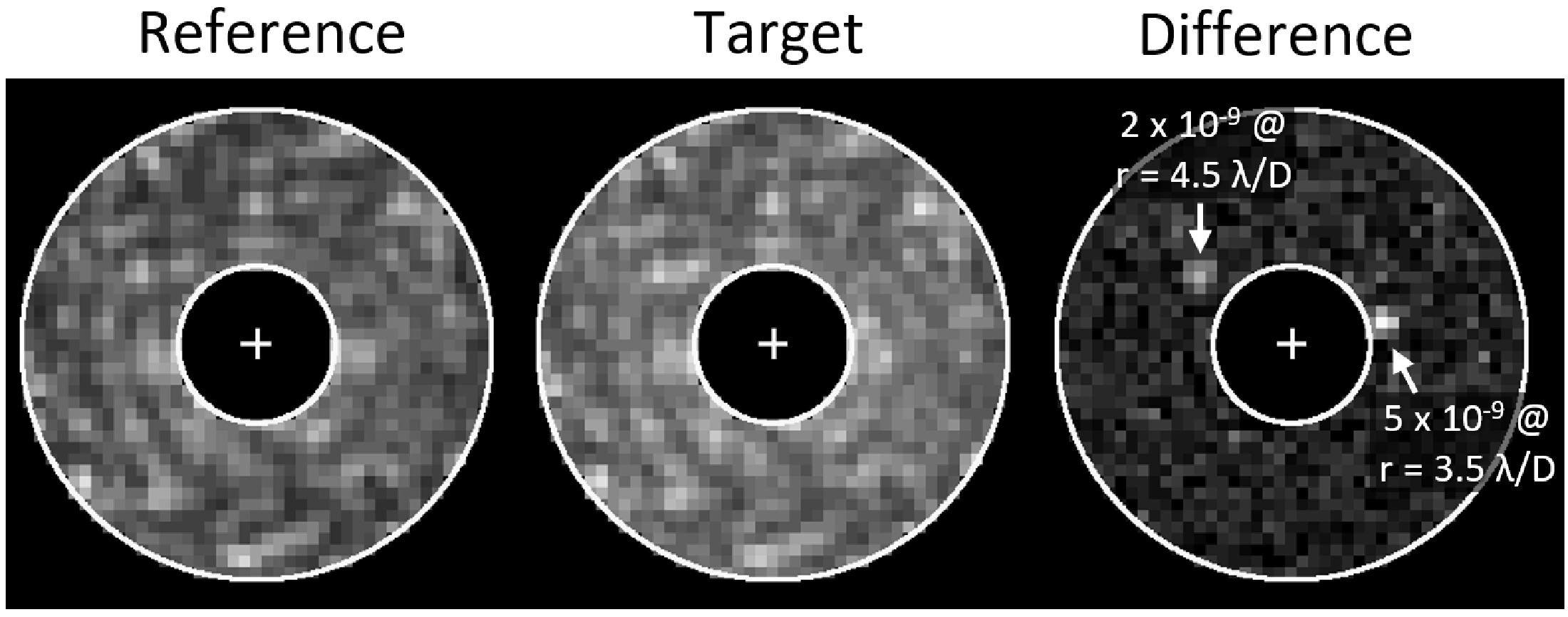}
\caption  
{ \label{fig:RDIOS11}
Results of the CGI OS11 integrated modeling run. The integrated modeling observing scenarios for CGI are labeled OS\#, where the \# is a serial number.  }
\end{figure}

In the RDI observing scenario, the telescope first points to the reference star, using which the coronagraph generates the dark hole, and creates the reference image. The telescope then slews to the target star to collect the science image. 
Fig.~\ref{fig:RDIOS11} shows the result of an integrated modeling run as an example. This run included both RDI and ADI, but, as explained in the previous section, ADI improvements are not included in the error budget. Hence the additional complexity of ADI processing is not mentioned in the remainder of this section.

The integration on each star can take hours to days, depending on its brightness, and is composed of many camera frames that are co-added to create the final image. The differential image is obtained by subtracting the scaled reference image from the target image:
\begin{equation}
\label{eq:RDI_difference}
\Delta I = I_{tar} - \beta\ I_{ref}
\end{equation}
The normalization factor $\beta$ accounts for the greater brightness of the reference star and the different integration time spent on it:
\begin{equation}
\label{eq:beta}
 \beta = \frac{F_\lambda^{t} }{F_\lambda^{r} }\cdot\frac{ t}{  t_r},
\end{equation}
where $F_\lambda^{t}$ and $F_\lambda^{r}$ are the spectral flux of the target and the reference stars, respectively, and $t$ and $t_r$ are the times of exposure to the target and the reference stars, respectively. In order to save time in generating the dark hole, reference stars are chosen from the ${\sim}100$ brightest stars in the sky. In the typical case, the reference star is ${\sim}3$ magnitudes brighter than the target. If the reference and target stars are of the same spectral type, this amounts to a factor of ${\sim}16$ in apparent brightness. Once the dark hole is generated, a final integration on the reference star produces the reference speckle image. The reference integration time, because of the brightness difference, would be shorter than what is needed to create the target star image. The time ratio can be optimized, and for the 3-mag difference a time ratio of ${\sim}5\times$ is close to optimum. Thus, for the typical case, with the $16\times$ brightness ratio and the $5\times$ integration time ratio, $\beta\approx0.32$.

The brightness ratio and the time ratio affect the contributions of noise from each source differently. 
The main sources of random noise are:
\begin{enumerate}
	\item $\sigma_{pl}$, the planet signal shot noise 
	\item $\sigma_{sp}$, speckle background shot noise
	\item $\sigma_{zo}$, zodiacal dust background shot noise (from both the local zodi, $\sigma_{lz}$, and exo-zodi, $\sigma_{ez}$) 
	\item $\sigma_{det}$, detector noise
\end{enumerate}
The first three error sources are photonic in origin while the last is electronic. 
The photonic errors require specific handling of the relevant throughput for each origin. For example, local zodi light is uniform within the field of view of the dark hole, while the planet signal is from a point source. The exo-zodi might be somewhere in between, from uniform to lumpy, and will require an explicit assumption about its brightness distribution. The throughput from a uniform source is larger than the signal point-source throughput by a factor of $1/\tau_{\rm PSF}$, where $\tau_{\rm PSF}$ is about 9\% for the HLC. 

The main source error which does {\em not} go down with integration time is $\sigma_{\Delta I}$, the residual speckle after differential imaging. For a perfectly stable telescope and coronagraph, and identical spectra for the target and reference star, the speckle pattern would be the same between the target and reference stars. In practice it does change, so that the RDI subtraction yields a residual level of speckle. 
From the standpoint of the effect of increasing science integration time, random error sources make contributions that go down with integration time, while thermal drifts usually go  {\em up}.
In the limit where the dominant change occurs in the switch between the two stars, and not by variations within each integration, subtraction yields a residual speckle, which increases with target integration time just as fast as the signal. In this case the residual speckle becomes the noise floor of the measurement. The residual speckle is subject to the coherent amplification effect described in Eq. \ref{eq:fov}, and requires careful treatment in the analytical model. This is detailed in a separate paper, but we will provide an overview of the treatment in Section~\ref{sect:Cstabilty}. 

Putting all the noise sources together, the total variance is given by:
\begin{equation}
\label{eq:totalVariance}
 \sigma_{tot}^2 = \sigma_{pl}^2 + \sigma_{sp}^2 + \sigma_{zo}^2 + \sigma_{det}^2 + \sigma_{\Delta I}^2  
\end{equation}
For the photometric random noise sources, the first four terms on the right, we can define corresponding ``variance rates'' and the total random noise variance rate $r_n$:
\begin{equation}
\label{eq:varianceRates}
 \sigma_{ph}^2 = r_n\ t = r_{pl}\ t + r_{sp}\ t + r_{zo}\ t + r_{det}\ t   
\end{equation}
For all but the detector term ($r_{det}t =\sigma^2_{det}$), these are also the count rates, because they have Poisson distributions (where the mean equals the variance). 

The last term on the right side of Eq.~\ref{eq:totalVariance} has a different dependence on target integration. As argued above, with $\sigma_{\Delta I}$, it is the standard deviation that grows linearly with time, so that:
\begin{equation}
\label{eq:sigmaDeltaI}
 \sigma_{\Delta I}^2 = r_{\Delta I}^2\ t^2\, .
\end{equation}
$r_{\Delta I}$ is called the residual speckle rate.  Section~\ref{sect:Cstabilty} describes how $ \sigma_{\Delta I}^2 $ is calculated. 

Each standard deviation that appears in Eq.~\ref{eq:totalVariance} refers to a contribution to the noise in the signal counts extracted from the differential image. To each of these, following Equations \ref{eq:xi_p} and \ref{eq:kappa}, there corresponds a flux ratio noise contribution:
\begin{equation}
\label{eq:FRNcorrespondence}
 \delta\xi_i = \kappa\cdot\sigma_i\, .
\end{equation}
In the error budget, the allocations are to the $\delta\xi_i$. Since independent errors $\sigma_i$ can be added in quadrature, so can the $\delta\xi_i$.

\subsection{SNR and time to SNR}

The photometric SNR, in terms of the count rates, can be obtained by first noting that all of the noise in the denominator of Eq.~\ref{eq:photometricSNR} is captured by $\sigma_{tot}$ (Eq.~\ref{eq:totalVariance}). 
With a regrouping of the noise variance that separates the random noise (Eq.~\ref{eq:varianceRates}) from the residual speckle noise (Eq.~\ref{eq:sigmaDeltaI}), the SNR can then be written as: 
\begin{equation}
\label{eq:SNR_rates}
{\rm SNR} = \frac{r_{pl}\ t}{\sqrt{r_n\ t + r_{\Delta I}^2\ t^2}}
\end{equation}
This equation shows the SNR achievable after integrating on the target star for a total time $t$. It also makes explicit the role of $r_{\Delta I}$ in the measurement noise floor.
Inverting this relation and solving for the target integration time, we arrive at the time-to-SNR equation:
\begin{equation}
\label{eq:tSNR}
t_{\rm SNR}  = \frac{{\rm SNR}^2\ r_n}{r_{pl}^2 - {\rm SNR}^2\ r_{\Delta I}^2 } 
\end{equation}
Implied but not appearing explicitly here is the additional time spent on the (brighter) reference star and the associated additional error. As noted above, the differential image normalization factor $\beta$ (Eq.~\ref{eq:beta}) depends on the reference star integration time and its brightness. We will shortly derive the full expression for $r_n$ and its dependence on $\beta$. 

In the time-to-SNR Equation (\ref{eq:tSNR}), the existence of a difference in the denominator points to the non-linear dependence of the integration time on SNR. If the speckle subtraction is not effective enough (i.e. $r_{\Delta I}$ is too high) or the desired SNR is too high for the available count rate from the planet, $r_{pl}$, the denominator can vanish or become negative, indicating no solution. Given this, it is useful to define, for a given observation, the {\em critical SNR}, where the denominator goes to zero and it takes infinite time to reach it:
\begin{equation}
\label{eq:crit_SNR}
{\rm SNR}_{crit}  = \frac{r_{pl}}{r_{\Delta I} } 
\end{equation}
This is the limiting SNR for this planet. A higher SNR is not achievable in any amount of time. Only a brighter planet, with a higher $r_{pl}$, can support a higher SNR. Alternatively, to see dimmer planets, we need to lower the contrast instability $r_{\Delta I}$.

\section{Photometric Errors}

The photometric errors will have contributions from both the target and reference stars. In this section we derive expressions for each of the photometric error contributors assuming negligible reference star contribution (i.e. infinite brightness reference star), then derive correction factors that account for the contribution from the reference star. 

\subsection{Detector Efficiency and Noise}
In high contrast exoplanet imaging, the photon rates from the planet can be extremely low ($10^{-2} s^{-1}$ or lower), with the result that in frame times shorter than a few hundred seconds, read noise will dominate. EMCCD's, like the Teledyne e2v CCD-311 used in CGI, provide amplification of the pixel signal with the use of a gain register. This feature of the EMCCD is an enabling technology for high contrast imaging and spectroscopy, but brings its own set of peculiarities that must be accurately represented in any performance model. We therefore will describe some of the salient features which pertain to our analytical model. 

\begin{figure}[ht]
\centering
\includegraphics[height=5cm]{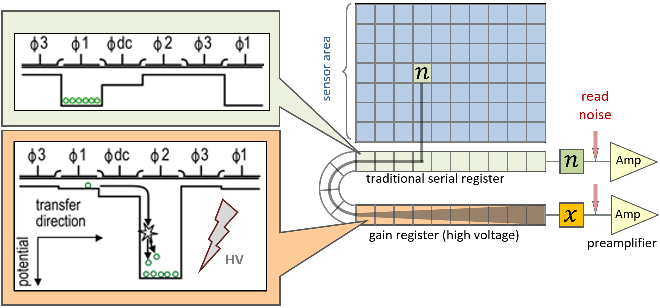}
\caption  
{ \label{fig:EMCCD}
The CGI uses EMCCD's manufactured by Teledyne e2v. These have two amplifiers, one for conventional mode operation, and one for EM gain. EM gain is effected by a clocking scheme that includes a high voltage step.  }
\end{figure}

The EMCCD's gain register is similar to a normal CCD's serial register, but also includes a high-voltage phase where the clock voltage amplitude can go into the 40 volt range. As each pixel's charge packet is clocked through the register (Fig.~\ref{fig:EMCCD}), the additional acceleration due to the high voltage increases the probability of impact ionization in collisions between charge packet electrons and silicon substrate ions. As a result, there is a non-negligible (order 1\%) probability that each electron in the charge packet can create an additional electron through impact ionization. This gain is adjustable by changing the high voltage amplitude. Since there are hundreds of gain stages in the register, the total gain can get very high. The CCD-311 has 604 gain stages. With a single stage probability of 1.42\% , for example, the mean electron multiplication (EM) gain is: $G  = (1+1.42\%)^{604}\approx 5000$.  

As in a normal CCD, the rows are first clocked in parallel into the standard serial register. Each row is then clocked serially through a corner register and then the gain register. An image area pixel that has collected $n$ electrons during frame integration will, on average, arrive at the sense node of the CCD pre-amplifier with $x=G\cdot n$ electrons. The readout of the sense node will include the sense node charge of $x$ plus the read noise. Thus the read noise, relative to the original image pixel counts of $n$ electrons, is diminished by a factor of $G$. This gain process is stochastic in nature, however. 
If $n$ electrons from a pixel enter the gain register, the probability distribution for $x$ electrons exiting the register is given by:\cite{Basden2003}
\begin{equation}
\label{eq:EMgainProb}
P(x,n) = \frac{x^{n-1}\ e^{-x/G}}{G^{n}\ (n-1)!} \, .
\end{equation}
The broadening effectively increases the noise associated with the measurement. It is customary to parameterize the added noise relative to the shot noise in $n$, and refer to it as the {\em excess noise factor} (ENF), associated with EM gain. Robbins et al.\cite{Robbins2003} calculate this to be:
\begin{equation}
\label{eq:ENF}
{\rm ENF}^2=2\ (G-1)\ G^{-(N+1)/N} + \frac{1}{G} \, ,
\end{equation}
where $G$ is the average register gain and $N$ is the number of gain stages. For the CCD-311, which has $N=604$ stages, ${\rm ENF}^2$ approaches its asymptotic value of 2 for all mean gain values above ${\sim}10$. The ENF is a multiplier on the noise or standard deviation. Thus, using gain effectively eliminates the read noise, but at the price of doubling all other sources of noise. 
\subsubsection{Photon Counting}
\label{photoncounting}
\begin{figure}[ht]
	\centering
	\includegraphics[height=4.5cm]{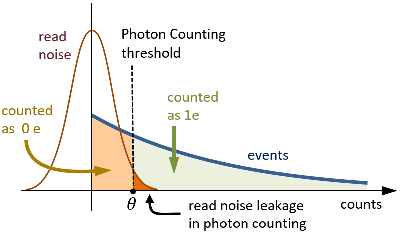}
	\caption 
	{ \label{fig:photonCounting}
		 Photon counting uses thresholding to separate read noise-only events from photonic events. In each frame pixels with counts less than a threshold $\theta$ are deemed to have zero electrons. The rest are deemed to have one electron. Depending on the threshold, there will be some amount of false positives from ``read noise leakge.''} 
\end{figure}

For the tech demo target star, which has a visual magnitude of 5, and even more so for most of the rest of the target stars, which have a median magnitude of ${\sim}6$, the scene is sufficiently dim to allow for a data reduction technique which eliminates the excess noise factor. The technique is called ``photon counting'' (PC).  In photon counting, the frame rate is increased to the point where the average counts per pixel are only ${\sim}0.1$ electrons per frame. Thus in any given frame, most of the pixels have zero counts. Most of those that have any photo-electrons have only a single electron, and a small fraction of those have 2 electrons. Photon counting involves setting a threshold that applies to the {\em post-gain} counts per pixel. The threshold is typically chosen at ${\sim}5$ times the read noise, 
\textcolor{revA}{
and all the pixels below the threshold are assigned 0 counts. The placement of the threshold is a trade between false positives (read noise leakage) and false negatives (thresholding inefficiency).}
This is illustrated in Fig.~\ref{fig:photonCounting}. The PC-processed frames are co-added to produce the final image. 

Photon counting allows for exquisitely low-noise imaging, but it adds two additional complications. The first is that thresholding has its own inefficiency. We can calculate the thresholding efficiency by noting that the post-gain probability distribution function for the single-electron pixels is given by Eq.~\ref{eq:EMgainProb}, with $n=1$. Integrating this function from the chosen threshold number of counts, $\theta$, to infinity gives the photon counting  efficiency:
\begin{equation}
\label{eq:PCefficiency}
\epsilon_{PC}=
\int^{\inf}_{\theta} P(x,1)\ dx = 
\int^{\inf}_{\theta} \frac{e^{-x/G}}{G}\ dx = e^{-\theta/G} \, .
\end{equation}
More precise estimation of $\epsilon_{PC}$ would have higher order corrections that account for $n=2,3,...$ etc.\cite{Nemati2020a}
The second effect that needs to be accounted for is the inherent inefficiency of assuming all non-empty pixels have precisely $n=1$ electrons. This effect is called {\rm coincidence loss} (CL).
\textcolor{revA}{Coincidence loss can be thought of as a loss of efficiency. It can be shown that the CL efficiency is given by\cite{Nemati2020a}:}
\begin{equation}
\label{eq:ConcidenceLoss}
\epsilon_{CL}=\frac{1-e^{-\Bar{n}_f}}{\Bar{n}_f}\, ,
\end{equation}
where $\Bar{n}_f$ is the mean number of counts per pixel per frame. If the frame rate is fast enough to allow $\Bar{n}_f\simeq\ 0.1\ {\rm e/pix/fr} $, the efficiency from this effect becomes $\epsilon_{CL}\simeq 95\%$.

These two effects can be simultaneously corrected to yield the true count rate. For example, consider $N_{fr}$ frames collected with an exposure time of $t_{fr}$ and an EM gain of $G$. If the frames are photon counted with a threshold of $\theta$, after co-adding the thresholded frames, the count rate for a pixel that accumulated a total of $N_{tot}$ counts (electrons) in the $N_{fr}$ frames is given by:
\begin{equation}
 	\label{eq:PCPC1pct}
 	r=\frac{-1}{t_{fr}}\ \cdot \ln\left( 1-\frac{N_{tot}/N_{fr}}{e^{-\theta/G}}\right)\, .
 \end{equation}
As described in (Nemati 2020)\cite{Nemati2020a} this first order correction is accurate to $\sim$1\%. In the same paper, the author also provides a solution with higher accuracy using Newton's method.

\subsubsection{Net Quantum Efficiency}\label{subsec:NetQuantum}
CCD manufacturer-supplied quantum efficiency (QE) data are limited primarily to the photon-to-electron conversion efficiency in the pixel, and the charge transfer efficiency in clocking. Charge transfer efficiency in parallel and serial clocking in CCD's is typically very high. 
The QE defined this way is mostly a function of the incident light wavelength, the reflectivity of the CCD surface, and the thickness of the substrate. Anti-reflection coatings can help optimize the QE for the bands that are used. 
For CGI's direct imaging operating mode, using the HLC coronagraph in Band 1 ($546-604~{\rm nm}$, Table~\ref{tab:CGIobsModes}), the QE is expected to be at least 85\%. But the net efficiency of the detector involves many more factors. Detailed description of the factors and their evaluation is beyond the scope of this paper, but we provide a brief overview here. 

Charge traps can cause a significant degradation of efficiency. The telescope in the L2 orbit experiences bombardment by galactic cosmic rays (GCR's) and solar protons. Displacement damage can occur when a high energy particle impacts a substrate ion in such a way that it is permanently removed or dislocated. The resulting perturbation to the local potential can then create a charge trap. \cite{Srour2003}  Solar flares can be times of rapid degradation for silicon devices in space.\cite{Feynman2002} There are different trap effects that can occur, and modeling the effect of charge traps requires knowledge of the occurrence rate of the different species.\cite{Nemati2016} It is also possible, using the technique of ``trap pumping'' to scan the device for not only the locations but also characteristics of existing charge traps.\cite{Hall2014a,Hall2014} Traps have a capacity, usually of one electron, and once they are filled they are no longer a factor until they emit the captured electron. This means that their effect {\em cannot} properly be described as a percentage loss, since the percentage depends on the incident photon flux. Overall, a sensor's charge transfer efficiency loss due to traps can best be modeled as a function of the number of electrons per pixel per frame. We use lab measurements of charge transfer loss, parameterized in this way, to include the effect of traps. The typical level of this efficiency, which we label $\epsilon_{\rm CTE}$, for direct imaging with the HLC coronagraph in band 1, is ${\sim}92\%$ after 21 months (the length of the official technology demonstration period for CGI) in the L2 orbit radiation environment.

Detector efficiency is also reduced due to more common effects like hot pixels. These are pixels with excessive dark current. They are identified and ''masked out'', meaning they are not used in the processing. As a result, they represent an inefficiency in signal detection. The efficiency from this effect, $ \epsilon_{\rm HOT}$, is ${\sim}98\%$ after 21 months in L2.

Another notable source of efficiency loss in an EMCCD operating in space is cosmic rays, which can deposit large amounts of energy in the CCD. These not only saturate the pixels but also create long tails along the serial direction. This effect has a mild dependence on EM gain, but in general has the effect of creating tails that are hundreds of pixels long. The effect of the tails can somewhat be mitigated in image processing, but since at this time those algorithms are not yet mature, in the analytical model we conservatively assume that the tail pixels are not usable in the frames and locations in which they occur. The efficiency factor from cosmic ray tails, $\epsilon_{CR}$, is ${\sim}99\%$.

Putting all of these effects together, the photon-counting net quantum efficiency of the detector, $\eta$, is 
\begin{equation}
\label{eq:netQE}
\eta = {\rm QE} \cdot \epsilon_{PC} \cdot \epsilon_{CL} \cdot \epsilon_{\rm CTE} \cdot \epsilon_{\rm HOT} \cdot \epsilon_{CR}
\end{equation}
For HLC in band 1, $\eta\simeq66\%$ after all the effects are included.

\subsubsection{Detector Noise}
\label{detectorNoiseSection}

The EMCCD can be operated with a wide range of gain settings, from 1 to over 7000, by changing the high-voltage clock amplitude and shape. The exact choice depends on the brightness of the source and the full well capacity of the detector, both in its image area and in the gain register. When the EMCCD image is used directly, the image counts must be divided by the a-priori known EM gain from previous calibration. The gain-corrected image is then in units of image area electrons. This mode of operation is called `analog' or proportional mode, and offers a reduction of the read noise by a factor of EM gain, $G$. Photon counting practically eliminates read noise for suitably chosen thresholds, but can only be used for the target stars. The reference stars are too bright for photon counting because of the high frame rates required. On the other hand, the brightness of the reference stars means they will be photon noise limited and detector noise is not a significant factor. 

After read noise, the remaining detector noise contributions are from 1) clock-induced charge and 2) dark current. 

Clock-induced charge (CIC) is present in all CCD's and is essentially the low-voltage equivalent of the EM gain process, except here it only produces noise. As the charge packets are clocked out, impact ionization causes the generation of CIC. Its magnitude depends on the clocking parameters, and is typically measured in units of electrons per transfer (from one pixel to the next). More typically, CIC is simply measured as an average, in units of electrons per pixel per frame (${\rm e/pix/fr}$). These are the same units as used for read noise, and since even in low-noise detectors the read noise $\sigma_{rd}$ is above ${\sim}1~{\rm e/pix/fr}$, while CIC is usually below $0.01~{\rm e/pix/fr}$, CIC is ``buried in the noise'' and is seldom considered in ordinary CCD applications. However, in an EMCCD the read noise is either greatly reduced (analog mode) or eliminated (photon counting mode), and CIC becomes noticeable. This is particularly true in photon counting, where the frame rates are fast, and CIC noise is incurred every frame. 

Dark current occurs when valence electrons in silicon are thermally excited into the conduction band of the lattice, and is hence highly temperature dependent. The CGI EXCAM will be operated at below $-85^\circ {\rm C}$ where the dark current is expected to be below $10^{-3}~{\rm e/pix/s}$. Dark current does not depend on the frame rate but only the total integration time. 

Detector noise scales with the number of pixels. Consistent with our treatment of the signal we assume aperture photometry in estimating the detector noise. The signal is contained in the core (see Eq.~\ref{eq:tau_core}) which subtends, on the sky, some solid angle $\Omega_{cor}(\lambda)$. Diffraction modeling of the optical system provides the PSF shape and size at each point in the dark hole. If the model PSF was evaluated at a different wavelength,  $\lambda_{m}$, the size of the core scales with $(\lambda/\lambda_{m})^2$.
The sampling of the PSF by the detector is an important consideration. The focal ratio of the EXCAM imaging lens is set to allow for Nyquist sampling of the PSF (i.e. 2 pixels per $\lambda/D$ where D is the diameter of the limiting aperture, usually the Lyot stop, projected to the entrance pupil) in the bluest of the wavelengths of interest. Labeling the critical/Nyquist-sampled wavelength, $\lambda_c$, the pixel planet scale is given by 
\begin{equation}
\label{eq:pixelPlateScale}
\theta_{pix}=\frac{1}{2}\cdot\left(\frac{\lambda_c}{D}\right) \, .
\end{equation}
Thus, for direct imaging, the number of detector pixels contributing to the signal and noise is estimated as:
\begin{equation}
\label{eq:mpix}
m_{pix} = 
\frac{\Omega_{cor}(\lambda)}{\theta_{pix}^2} = 
\Omega_{cor}(\lambda_m) \cdot \left(\frac{ \lambda }{ \lambda_m }\right)^2 \cdot \left(\frac{2D}{\lambda_c}\right)^2 \, .
\end{equation}
In the narrow FOV HLC planet imaging mode (Table~\ref{tab:CGIobsModes}) mpix works out to be $4.95$ pixels. We will derive the appropriate expression for the spectroscopy case later in Section~\ref{sect:spectroscopy}.

The total noise variance from the detector, operating in ``analog'' (ANL) mode, contributing to the planet photometry error can now be written as:
\begin{equation}
\label{eq:ANLdetectorNoise}
\sigma^2_{\rm ANL}={\rm ENF}^2\ \left[  m_{pix}\ i_d\ t + m_{pix}\  {\rm CIC}\ \left( \frac{t}{t_{f}} \right) \right] + 
m_{pix}\ \left( \frac{\sigma_{rd}}{G}  \right)^2 \ \left( \frac{t}{t_{f}} \right) \, ,
\end{equation}
were ENF is the excess noise factor (given by Eq.~\ref{eq:ENF}), $i_d$ is the dark current in ${\rm e/pix/s}$, CIC is in ${\rm e/pix/fr}$, $t$ is the total integration time, and $t_f$ is the frame exposure time. This is the noise for the analog mode operation, which is used in creating the dark hole and also in the CGI low-order wavefront sensor (LOWFS) camera, called LOCAM. In these cases the fluxes are too high for photon counting since photon counting needs fast enough frame rates to make $\Bar{n}_f=0.1$ (Eq.~\ref{eq:ConcidenceLoss}).

\textcolor{revA}{
In photon counting (PC) mode, ${\rm ENF}$ has a value of 1, and the read noise term 
effectively vanishes. We say ``effectively'' because, depending on the photon counting  threshold $\theta$, some read noise ``leakage'' occurs when the Gaussian tail from the read noise distribution crosses the threshold (Fig.~\ref{fig:photonCounting}). Since we incur this error once per pixel per frame, the effective count rate from this leakage can be written as:
\begin{equation}
\label{eq:PCreadNoiseLeakage}
r_{lk}=\frac{1}{2}\ {\rm erfc} \left( \frac{\theta/\sigma_{rd}}{\sqrt{2}} \right)
\cdot \frac{1}{t_f}
\, ,
\end{equation}
where ${\rm erfc}$ is the complementary error function. 
}

\textcolor{revA}{
The photon counting detector variance, $\sigma^2_{PC}$, does not include ENF but does retain a read noise leakage term:
\begin{equation}
\label{eq:PCdetectorNoise}
\sigma^2_{PC}=  m_{pix} \left(
i_d\ +  r_{lk}\  +  \frac{\rm CIC}{t_f}\right)  t  \, .
\end{equation}
}

\textcolor{revA}{
In what follows, we will generalize the detector variance to $\sigma^2_{det}$, but keep in mind that the observing scenario calls for photon counting so $\sigma^2_{det}=\sigma^2_{PC}$.
}

\subsection{Planet Shot Noise}

The next few noise terms are photonic in origin, and all contribute as shot noise. Since shot noise follows Poisson statistics, the variances for these noise terms are simply equal to the respective means.

Following Eq.~\ref{eq:r_pl}, the planet shot noise variance is given by: 
\begin{equation}
\label{eq:planetShot}
\sigma^2_{pl} =    {\rm ENF}^2\ \cdot F_{\lambda}\ \Delta\lambda\ \xi_{pl}\ A_{col}\ \tau_{pl}\ \eta\ t \, .
\end{equation}
\textcolor{revA}{
Here and below, it should be kept in mind that ENF is simply 1 for photon counting, which is how the observation will be processed. For analog observations, such as very bright targets, photon counting might become unfeasible, and ENF would be included, given by Eq.~\ref{eq:ENF}.
}

\subsection{Speckle Shot Noise}

The mean speckle rate at the planet signal location is proportional to the coronagraph starlight leakage to that point. In Eqn.~\ref{eq:contrast} and the preceding paragraph, this leakage was expressed as a leakage throughput $\tau_{\ell}(u,v)$, namely photons per modeling pixel as a fraction of photons incident on the collecting area. Expressed in terms of contrast, this throughput is given by $\tau_{\ell}=C\ \tau_{pk}$, following  Eqn.~\ref{eq:contrast}, where we have dropped the explicit notation of the coordinates $(u,v)$ .

\textcolor{revA}{
The CGI coronagraph model\cite{Krist2015b} computes the contrast and related quantities with $\Omega_m$ corresponding to a modeling pixel with finer resolution than the detector.}
$\tau_{\ell}$ is the leakage fraction into {\em a single} modeling pixel, so the light arriving within the entire core region has a higher throughput by a factor of:
\begin{equation}
\label{eq:m_cor}
m_{cor} =    \frac{\Omega_{cor}}{\Omega_{m}}  \,  ,
\end{equation}
where $ \Omega_{cor} $ is the solid angle subtended by the core region, projected on the sky, while $\Omega_{m}$ is the solid angle associated with a single image area pixel in the diffraction model that computes the contrast and throughput. $m_{cor}$ is the analog of $m_{pix}$ in Eq.~\ref{eq:mpix}, except that it refers to the modeling pixels. 

With the foregoing in mind, the mean speckle rate is given by:
\begin{equation}
\label{eq:r_sp}
r_{sp} =  F_{\lambda}\ \Delta\lambda\ C\ \tau_{pk}\ m_{cor} \ A_{col}\ \tau_{sp}\ \eta\ \, ,
\end{equation}
where $\tau_{sp}$ is different from $\tau_{pl}$ (Eq.~\ref{eq:planetThroughput}) only by not needing the core throughput factor: 
\begin{equation}
\label{eq:tau_sp}
 \tau_{sp} =  \tau_{rt} \cdot \tau_{fil} \cdot \tau_{pol} \, .
\end{equation}

The mean speckle background after integrating over time $t$ is simply $r_{sp}\ t$. The speckle shot noise variance is given by:
\begin{equation}
\label{eq:speckleShot}
\sigma^2_{sp} =   {\rm ENF}^2\ \cdot F_{\lambda}\ \Delta\lambda\ C\ \tau_{pk}\ m_{cor} \ A_{col}\ \tau_{sp}\ \eta\ t \, .
\end{equation}

\subsection{Zodi Shot Noise}

Light from the zodiacal clouds, both in our solar system and around the target star, forms an  extended background in the direct image. Zodi background is characterized by its surface brightness, $\zeta$, measured in magnitudes per square arcsecond (${\rm mag}/as^2$). 
Observationally, if within a measurement area $\Omega_{ob}$ an integrated magnitude of $m$ is observed, then the surface brightness is given by:
\begin{equation}
\label{eq:surfaceBrightness}
\zeta =   m + 2.5\ \log_{10}(\Omega_{ob}/as^2) \, .
\end{equation}

The surface brightness from the local zodi is expected to be $\zeta_\Sun \simeq 23\ {\rm mag}/as^2$.\cite{Stark2014b}
The surface flux (in photons per second per square arcsecond) from local zodi is thus given by:
\begin{equation}
\label{eq:zodiSpectralFlux}
\frac{\partial \Phi_\Sun}{\partial \Omega}  =   F^0_\Sun\ \Delta\lambda \cdot  \frac{10^{-0.4\ \zeta_\Sun}}{as^2}  \, ,
\end{equation}
where $F^0_\Sun\ \Delta\lambda $ is the integrated 0-magnitude flux (photons per second), assuming the solar spectrum, within the band of interest. Thus the local zodi ($lz$) rate into a PSF-core sized region of the sky is given by:
\begin{equation}
\label{eq:localZodiRate}
r_{lz} = \frac{\partial \Phi_\Sun}{\partial \Omega}  \ \Omega_{cor} \ A_{col}\ \tau_{lz}\ \eta \, ,
\end{equation}
where, as before, $\Omega_{cor}$ is the PSF core solid angle, and $\tau_{lz}$ is different from the planet throughput $\tau_{pl}$ (Eq.~\ref{eq:planetThroughput}) by the core encircled energy factor ($\tau_{PSF}$,  cf. Eq.~\ref{eq:tau_core}). This is 
because the planet is a point source, and only a fraction of the energy ends up in the core, while the local zodi is uniform in brightness over the field of view and $\tau_{\rm PSF}$ so that, by symmetry, this loss is not present:
\begin{equation}
\label{eq:tau_localZodi}
 \tau_{lz} = \tau_{occ} \cdot \tau_{rt} \cdot \tau_{fil} \cdot \tau_{pol} \, .
\end{equation}

The local zodi shot noise variance over an integration time $t$ is given by:
\begin{equation}
\label{eq:localZodiShot}
\sigma^2_{lz} =   {\rm ENF}^2\  r_{lz}\ t \, .
\end{equation}

In the case of exo-zodi, uniformity over the region around the planet core is not guaranteed, and there may be a throughput loss for this background. The amount of the loss depends on the zodi structure assumed for the exo system, and modeling is required to estimate the loss for the given structure.  In general, the exo-zodi throughput, $\tau_{ez}$, will be somewhere between the limiting cases of $\tau_{lz}$ and $\tau_{pl}$. 

Following an approach similar to Stark et al.~\cite{Stark2014b}, we assume that a ``1-zodi'' surface brightness exo-zodi system has the same surface brightness as our local zodi when viewed from outside, and when the spectrum of the star is the same as the sun. If our solar system is viewed from the outside, the full depth of disk would contribute to the surface brightness, so that instead of  23 ${\rm mag}/as^2$ 
we would see  $\zeta_* \simeq 22\ {\rm mag}/as^2$. 
Thus for the exo zodi spectral flux, we modify Eq.~\ref{eq:zodiSpectralFlux} by using the spectrum of the host star, 
accounting for luminosity difference, the separation of the planet in AU, 
and the density of the exo system dust at this separation relative to ours (i.e. the number of zodi's, $n_{zo}$):
\begin{equation}
\label{eq:exozodiSpectralFlux}
\frac{\partial \Phi*}{\partial \Omega}  =   n_{zo} \cdot
F^0_*\ \Delta\lambda \cdot  \frac{10^{-0.4\ \zeta_*}}{as^2} \ 10^{-0.4\cdot(M_*-M_\Sun)} \ \left( \frac{1\ {\rm AU} }{a_{p}} \right)^2 \, ,
\end{equation}
where $M_*$ is the absolute magnitude of the host star, $M_\Sun$ is the absolute magnitude of the Sun (4.83 mag in V band), and $a_{p}$ is the planet orbit radius (in AU).

Thus the local zodi ($lz$) rate into a PSF-core sized region of the sky is given by:
\begin{equation}
\label{eq:exoZodiRate}
r_{ez} = \frac{\partial \Phi_*}{\partial \Omega}  \ \Omega_{cor} \ A_{col}\ \tau_{ez}\ \eta \, ,
\end{equation}
The exo zodi shot noise variance over an integration time $t$, which we label as $\sigma^2_{ez}$, is given similarly to Eq.~\ref{eq:localZodiShot}, only with labels changed accordingly. 

\subsection{Reference Star Noise Contribution}
\label{sect:RDInoise}
So far we have only considered the random noise contribution from the target star. We now derive the enhancement factor for each random noise term coming from the reference star. All these factors tend to unity (no enhancement of noise) as the reference star becomes brighter. The spatial context in each case will be the PSF core-sized region around the planet location on the dark hole. The temporal context is the differential image of Eq.~\ref{eq:RDI_difference} except now we restate it in terms of the counts involved:
\begin{equation}
\label{eq:RDIcounts}
\Delta n = n_{tar} - \beta\ n_{ref} \, ,
\end{equation}
that is, within a core-sized region at the planet location, the integration on the target star (during time $t$) yielded $n_{tar}$ counts, while the integration on the reference star (during time $t_r$) yielded $n_{ref}$ counts. The reference counts are normalized by the factor $\beta$ given by Eq.~\ref{eq:beta} before the subtraction. Since the reference star only has the background terms,  there is obviously no enhancement factor for the planet noise. For the discussion in this section it is convenient to think of the target star also not having a planet, so the expectation value of the difference is $\langle \Delta n\rangle=0$ and hence 
\begin{equation}
\label{eq:ntar_nref_noplanet}
\quad \quad\quad\quad n_{tar} = \beta\ n_{ref}\, .   \quad \text{(no planet)} 
\end{equation}
The differential image total noise (Eq.~\ref{eq:totalVariance}) can also be written as the sum of the contributions from the target and reference star:
\begin{equation}
\label{eq:totNoise2}
\sigma^2_{tot} = \sigma^2_{tar} + \sigma^2_{ref} \, .
\end{equation}

\subsubsection{Speckle background}

For the speckle noise, we assume this is the only background source, so that $n_{tar}$ consists only of speckle counts, and similarly for $n_{ref}$. Since shot noise follows Poisson statistics, the mean is equal to the variance: $n = \sigma^2$. Hence, Eq.~\ref{eq:ntar_nref_noplanet} implies that:
\begin{equation}
\label{eq:refnoise_speckle}
\sigma_{ref}^{sp}=\sqrt{\beta}\ \sigma_{tar}^{sp} \, .
\end{equation}
 
\subsubsection{Local zodi background}

The reference star may be 20 degrees away from the target star, but the reference star is chosen such that the change in the solar angle  is as small as possible, in order to maintain the same thermal loads on the observatory. The solar angle is the angle between the line of sight and the vector to the Sun (Fig.~\ref{fig:RDI_ADI}). As a result of this, it is expected that the surface brightness of the local zodi will not change appreciably between the two star observations.
The background count rate from local zodi, therefore, is the same between target and reference. The shot noise therefore is only different because of the different integration times on target ($t$) and reference ($t_r$). The reference counts from local zodi are smaller than the target counts by a factor of $t_r/t$ so that the shot noise is smaller by $\sqrt{t_r/t}$. The normalization factor from differential imaging, $\beta$, which is used to match the mean speckle rate between the two observations, now linearly affects the noise from the reference star. When the normalization factor  is included, the net effect for local zodi (lz) shot noise is: 
\begin{equation}
\label{eq:refnoise_localzodi}
\sigma_{ref}^{lz}=\beta\ \sqrt{ t_r / t}\ \sigma_{tar}^{lz} \, .
\end{equation}

\subsubsection{Exo zodi background}

For exo-zodi, the densities of the disks may be different for the target and reference stars, and of course the stars have different brightness. If we assume that the densities are the same, then the difference would only be from the relative brightness of the two stars. In this case the factor is identical to the factor for speckles, along the same lines of reasoning, so that the net effect for exo zodi (ez) is: 
\begin{equation}
\label{eq:refnoise_exoZodi}
\sigma_{ref}^{ez}=\sqrt{\beta}\ \sigma_{tar}^{ez} \, .
\end{equation}
 
\subsubsection{Detector noise}

Neglecting read noise, the important remaining detector noise sources are dark current and clock induced charge. Assuming identical operating conditions between the science and reference stars, the dark current ($dc$) counts from the reference star are  $n^{dc}_{ref}=(i_d m_{pix} )\ t_r$ so that $n^{dc}_{ref}=(t_r / t)\  n^{dc}_{tar}$. Dark current noise is also a shot noise, variance equal to the mean. When the normalization factor from differential imaging, $\beta$, is included, the reference variance is $\sigma_{ref}=\beta\sqrt{t_r/t}\ \sigma_{tar}$. The detector clock-induced charge contribution from the reference star is related to the contribution from the target star in the same way as the dark current, by simple inspection of the form of these two error sources in Eq.~\ref{eq:PCdetectorNoise}. The full detector error is the quadrature sum of the dark current and CIC contributions, so the detector (det) part of the reference noise is related to the target noise according to: 
\begin{equation}
\label{eq:refnoise_det}
\sigma_{ref}^{det}=\beta\ \sqrt{ t_r / t}\ \sigma_{tar}^{det}  \, .
\end{equation}
Read noise is not a factor since in an EMCCD the read noise is effectively zero. However, had read noise been important, the ratio would be the same as CIC and dark current. 

The factors are summarized in Table~\ref{tab:refStarFactors} for the five error categories that apply. There are two ways one can book-keep the additional error from the reference star: one can either add up all the reference star contributions in quadrature into one total number, to be added in quadrature itself with the target star contribution, or apply an error enhancement factor ($k_x$) to each of the error sources to account for the differencing. The bottom row gives the variance enhancement factors $k_x$ for each error source. For the speckle, for example, the table is to be read to give 
$k_{sp}=1+\beta$.

\begin{table}[ht]
\caption{Reference star random noise contribution relative to that of the target star. The second row gives the multiplied $k_x$ that must be applied to a calculated target random noise term to account also for the reference star observation.} 
\label{tab:refStarFactors}
\begin{center}       
\begin{tabular}{r  c c c c} 
         &  speckle ($k_{sp}$)  & local zodi ($k_{lz}$) & exo zodi ($k_{ez}$) & detector ($k_{det}$) \\ \hline\hline
\rule[-1ex]{0pt}{3.5ex} $\sigma_{ref}/\sigma_{tar}$ &  $\sqrt{\beta}$ & $\beta\ \sqrt{ t_r / t}$ & $\sqrt{\beta}$ & $\beta\ \sqrt{ t_r / t}$ \\
\rule[-1ex]{0pt}{3.5ex} $k_x=\sigma^2_{tot}/\sigma^2_{tar}$ &  $1+\beta$ & $1+\beta^2\ t_r/t$  &  $1+\beta$ & $1+\beta^2\ t_r/t$ \\
\hline 
\end{tabular}
\end{center}
\end{table} 

The the total noise variance rate, introduced in Eq.~\ref{eq:varianceRates}, can now be augmented to include also the contribution from the reference star observation in RDI observing by using the enhancement factors defined in Table~\ref{tab:refStarFactors}:
\begin{equation}
\label{eq:refnoise_total}
r_n = r_{pl} + k_{sp} r_{sp} + k_{lz} r_{lz} + k_{ez} r_{ez} + k_{det} r_{det}  \, .
\end{equation}
where we have separated the zodi term into the local and exo contributions since they have different noise enhancement factors.  

\subsection{Modifications to the Calculations for Spectroscopy}
\label{sect:spectroscopy} 
In spectroscopy mode, Roman CGI will use a prism-slit spectrograph with a required resolving power of $R = \lambda/\delta\lambda=50$ at 730~nm. On a planet with flux ratio $5\cdot10^{-8}$, the CGI must achieve ${\rm SNR}=10$ per spectral element within a 15\% wavelength band. This is essentially a photometric SNR requirement, but per spectral element. As such, much of what has preceded remains applicable, with a few relatively simple modifications. 

The planet PSF falls on a slit which is sized to let the core go through at the central wavelength. The core light is dispersed by the prism and imaged onto the detector. Spectral elements are identified from the detector image using prior calibration data. The number of spectral elements is $N_{sp} = R\cdot {\rm BW}$, where ${\rm BW}$ is the wavelength bandwidth as a fraction of the central wavelength. 
Considering a representative spectral element, for example the central one, a fraction of the core light is contained within it. Moreover, on the detector, it subtends a given number of pixels. We now address each of these two factors. 

We define the ``signal region core fraction'', $f_{SR}$ as the fraction of the light in the core that is within this one spectral element. Assuming a flat spectrum, $f_{SR}$ would be approximately $1/N_{sp}$, while for direct imaging and aperture photometry, where we use all of the core, $f_{SR}=1$.
We can thus generalize Eq.~\ref{eq:r_pl} to both imaging and spectroscopy by including the signal region core fraction:
\begin{equation}
\label{eq:r_pl_fSR}
 r_{pl}=f_{SR}\ F_{\lambda}\ \Delta\lambda\ \xi_{pl}\ A_{col}\ \tau_{pl}\ \eta \, ,
\end{equation}
The same thing can be done for the speckle rate (Eq.~\ref{eq:r_sp}) and local and exo zodi rates (Eqs.~\ref{eq:localZodiRate}, \ref{eq:exoZodiRate}). The detector terms in 

In Eq.~\ref{eq:mpix} we labeled the number of detector pixels affecting the SNR as $m_{pix}$ and gave an expression for calculating it for the direct imaging case. For the prism spectrograph case, the region of interest comprises the detector pixels in a single spectral element. In the spectral or dispersion direction, the number of pixels per spectral element is given by 
\begin{equation}
\label{eq:n_pix_Rlimited}
 n_{pix}^{sp} =  \frac{d\theta}{d\lambda}\  \frac{f_c}{x_d}\ \delta\lambda_R \, ,
\end{equation}
where $d\theta/d\lambda$ is the spectrograph angular dispersion (a function of $\lambda$), $f_c$ is the camera focal length, $x_d$ is the width of a camera pixel, and $\delta\lambda_R$ is the spectral element width in wavelength units: $\delta\lambda_R = \lambda/R)$. The focal length $f_c$ is constrained by the requirement to Nyquist-sample the PSF core by the detector pixel: $f_c=2x_d D_c/\lambda$, where $D_c$ is the compressed beam diameter at the pupil before the imaging lens. 

In the {\em cross-spectral} (``$xs$'') dimension, the number of SNR-relevant pixels is:
\begin{equation}
\label{eq:n_pix_crossSpec}
n_{pix}^{xs} =  w_{\rm PSF}^{xs}\  \frac{\lambda}{D_c}\ \frac{f_c}{x_d}  \, ,
\end{equation}
where $ w_{\rm PSF}^{xs}$ is the width of the PSF core in the cross-spectral direction. 

The product of the number of pixels in the two directions gives the total number detector pixels for one spectral element:
\begin{equation}
\label{eq:mpix_spectrscopy}
m_{pix}= n_{pix}^{sp}\ n_{pix}^{xs} = \frac{d\theta}{d\lambda}\ \delta\lambda_R\ \left( \frac{f_c}{x_d}\ \right)^2  w_{\rm PSF}^{xs}\  \frac{\lambda}{D_c}  \, .
\end{equation}
For the planet spectroscopy mode (see Table~\ref{tab:CGIobsModes}), $m_{pix}$ works out to be about \textcolor{revA}{$17.0$ pixels.}

In summary, the modifications for the spectroscopy mode are to:
\begin{enumerate}
    \item insert a factor $f_{SR}$ into the planet signal rate, as done in Eq.~\ref{eq:r_pl_fSR};
    \item do the same thing for all the shot noise rates in Eq.~\ref{eq:refnoise_total};
    \item replace, in the last term in Eq.~\ref{eq:refnoise_total} (with $r_{det}$), the $m_{pix}$ expression for spectroscopy, Eq.~\ref{eq:mpix_spectrscopy}, instead of the one for direct imaging, Eq.~\ref{eq:mpix}.
\end{enumerate}

\section{Contrast and Speckle Instability}
\label{sect:Cstabilty}  
One complete branch of the error budget is related to contrast and its instability. Temporal geometrical changes in the optical system, whether rigid body movements or structural deformations of any of the optics, from the telescope to the coronagraph, and many other subtle effects, can change the dark hole field and hence the speckle pattern. These temporal changes undercut the effectiveness of RDI in accounting for the speckles. Also, unlike the photometric errors described in the previous section, longer integration times do not reduce these errors. If the speckle pattern changes between the reference and target observations, the RDI residual speckle's contribution to the noise grows as fast as the signal.

Since the change in the speckle involves the diffraction of the complex field, a detailed treatment of the problem requires a full diffraction model. We have developed such a model, called PROPER, and have described it in detail elsewhere.\cite{Krist2015b}
For the purposes of performance modeling and error budgeting, however, a faster analytical approach is desired. Description of the treatment and its justification requires a significant space. A more detailed description is being prepared by one of the authors (BK) for a separate paper in which the premises and assumptions, the derivation of the results, and the details of its implementation will be fully presented. Here we will lay out the general approach and some of the key results.

\subsection{The RDI Ensemble} 

We consider a dark hole (DH) and one core-sized region of interest. Light that falls on the dark hole can be from incoherent phenomena, such as from background galaxies or scatter from the solar system objects, or it can be coherent light, coming from the target star, scattering or ``leaking'' into the dark hole. The differential image is a difference of intensities, which are in turn determined by the fields. Focusing on coherent effects only, the electromagnetic field, which we treat as a scalar phasor, is time varying due to disturbances, and there is, after differential imaging, a residual intensity map (i.e a residual image). This residual intensity map, as we will shortly see, is a function of the original field and its temporal variation.

\begin{figure}[ht]
\centering
\includegraphics[height=4cm]{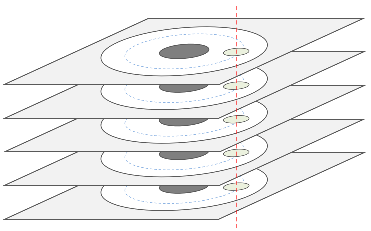}
\caption 
{ \label{fig:ensembleROI}
An stack of RDI images from an ensemble of similar visits to a specific target star. Each image is the ``final'' subtracted image, from which a planet signal could be detected. For each,  the dark hole with a core-sized region of interest (ROI), is denoted. The statistical ensemble comprises the ROI’s from these visits.} 
\end{figure} 

Consider an ensemble of differential imaging instances (Fig.~\ref{fig:ensembleROI}), where in each instance we have formed a \textit{differential} (RDI) image after chopping back and forth between the target and reference stars and subtracting the co-added, scaled reference image from the co-added target image (Eq.~\ref{eq:RDI_difference}). Across this stack of imaging instances, we select a region of interest (ROI), typically PSF core-sized and at a radius where we expect the planet signal to be. For  error budgeting, we desire a statistical description, in particular the ensemble-wide variance, of the total intensity within the ROI across these differential images. 

\begin{figure}[ht]
\centering
\includegraphics[height=4.5cm]{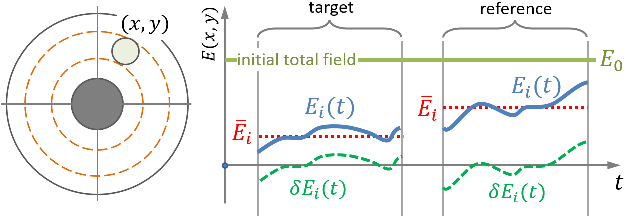}
\caption {Considering an ROI centered at some location $(x,y)$ in the dark hole (left), we consider an intial total field $E_0$ and varios contributing fields $E_i(t)$. Each of these is then broken down into its mean $\bar{E}_i$ and its variation $\delta E_i(t)$ during target star and the reference star observations (right).} 
\label{fig:FieldTimeEvolution}
\end{figure}

Taking into account for now only coherent phenomena, the total field in the ROI at any moment is the sum of many coherent contributions with fields adding directly. The focal plane field at any given time is therefore the sum of an initial, disturbance-free field $E_0$ (a scalar complex quantity in our treatment), and a number of time-varying additional contributions from various disturbance sources (e.g. thermal variations) or modes (e.g. a Zernike mode) . If only contribution $i$ is active, Fig.~\ref{fig:FieldTimeEvolution} illustrates the quantities of interest.

During the stare at the target, the time-varying field from the $i^{th}$ contributor $E_i(t)$ can be considered as the sum of its mean $\bar{E}_{i}$, and its variation $\delta E_i(t)$:
\begin{equation}
    E_i(t) = \bar{E}_{i} + \delta E_i(t) \, .
    \label{eq:Ei}
\end{equation}

\textcolor{revA}{
Detectors measure image plane intensity, given by the modulus-squared of the (complex) field. From the $i^{th}$ contributor, the detector receives the intensity:
\begin{equation} 
\label{eq:imagePlaneIntensity}
I_i(t) = |E_i(t)|^2 = |\bar{E}_i|^2 + |\delta E_i(t)|^2 + 
        2\Re\{\bar{E}_i^*\delta{E_i(t)}  \} \, ,
\end{equation}
where $\Re$ and $\Im$ are represent the real and imaginary parts of a complex number, respectively. 
This is the contribution to the intensity from source $i$. 
We write the total field  $E$ at $(x,y)$ as the sum of a disturbance-free initial field $E_0$ and the contributions from the various disturbances: 
\begin{equation} 
\label{eq:totalField}
E = E_0 + \sum_{i=1}^{n}E_i(t) =  \sum_{i=0}^{n}E_i(t) \, ,
\end{equation}
$E_0$ is strictly constant, while $E_i\ (i>0)$ are time varying contributions.
}
\textcolor{revA}{
\subsection{Total Intensity from All the Field Contributions}
}

\begin{figure}[ht]
    \centering
    \includegraphics[height=6.5cm]{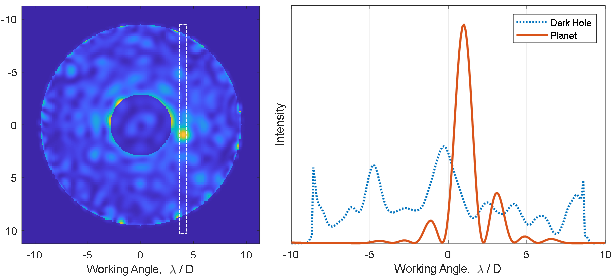}
    \caption{Example of a differential image $\Delta I$ (random noise not included) for a single visit (see Fig.~\ref{fig:ensembleROI}). A fit for the signal, informed by the known planet PSF shape, tends to lessen the speckle background contribution.}
    \label{fig:PSFvsDHshape}
\end{figure}

\textcolor{revA}{
For the full set of field contributions $\{E_i\}$, with each broken down into a mean and a variation per Eq.~\ref{eq:Ei}, the total intensity is given by:
\begin{equation}
    I(t) =\left| E(t)\right|^2 = \left|\sum_{i=0}^{n}E_i(t)\right|^2
    = \left|\sum_{i=0}^{n} (\bar{E}_{i} + \delta E_i(t))\right|^2 \, .
    \label{eq:totintensity}
\end{equation}
If we make the assumption that the mean fields and the mean field variations are uncorrelated among the different contributors, it can be shown that the average intensity during an observation will be given by: 
\begin{equation}
    I_{av}\equiv \langle I(t)\rangle =
           \sum_{i=0}^{n}\lvert\bar{E}_i\rvert^2  + 
    \langle\sum_{i=0}^{n} \lvert \delta E_i(t)\rvert^2\rangle \, .
\label{eq:Iav}
\end{equation}
The first sum is of the average intensity contributions, while the second sum is of the field variance contributions. On the right hand side, in the first sum we have dropped the $\langle\ldots \rangle$ temporal average since the quantities being averaged are already time averages and constant over the time interval. Bearing in mind that we are here dealing with means and variances, we define $M_i\equiv\bar{E}_i$ and $M_{tot}\equiv\sum\bar{E}_i$. 
Again one can show that with uncorrelated error sources, the cross terms will vanish in the temporal average so that:
\begin{equation}
    \lvert M_{tot} \rvert^2 = 
    \sum_{i=0}^{n}\lvert\bar{E}_i\rvert^2 \, .
\label{eq:MtotEi}
\end{equation}
Thus for each observation, the average intensity is the sum of the total mean field $M_{tot}$, modulus squared, and the total variance $V_{tot}$:
\begin{equation}
    I_{av} = \lvert M_{tot} \rvert^2 + V_{tot} \, ,
\label{eq:IavMtotVtot}
\end{equation}
where $V_{tot} = \sum_i V_i$ is the total variance and 
$V_i=\langle\lvert\delta E_i(t)\rvert^2\rangle$
is the contribution from one disturbance source.
}
\textcolor{revA}{
\subsection{Variance of the Differential Image}
}
\textcolor{revA}{
In a differential image, we subtract $I_{av}^{ref}$ (scaled for brightness) from $I_{av}^{tar}$ (Eq.~\ref{eq:RDI_difference}). We can ask, considering a core-sized (i.e. $\sim\lambda/D$ sized) ROI in the dark hole, and an ensemble of such differential images, what is the variance of the differential intensity among the ensemble (Fig.~\ref{fig:ensembleROI}). We can think of the steps as
\bigbreak
    Individual Frames $\rightarrow$ Differential Image $\rightarrow$ Post-processed Image $\rightarrow$ Extracted Signal
\bigbreak
In signal extraction, the fit for the signal will help in suppressing the background due to shape mismatch. In Fig.~\ref{fig:PSFvsDHshape}, the dark hole and planet can be seen, as well as a slice through the image. The planet PSF shape is well known, but pixelation and noise will in general reduce the effectiveness of the fit in rejecting backgrounds. The unrejected background will cause some overcount or undercount of the signal. This is an important quantity:
\textit{The ensemble standard deviation of the miscounts is the measure of the error from speckle instability, an error which we call speckle noise. }
Speckle noise constitutes the largest branch of the error budget. 
}

\textcolor{revA}{
Keeping in mind that $M_{tot}$ in Eq.~\ref{eq:IavMtotVtot} is complex, 
the differential intensity in terms of sensitivities is given by:
\begin{equation}
\label{eq:Dif_Sum_I}
    \Delta I=\sum_{i=0}^{n}\left[\frac{\partial I_{av}}{\partial V_{i}}\Delta V_{i}+\frac{\partial I_{av}}{\partial \Re(M_{i})}\Delta \Re(M_{i})+\frac{\partial I_{av}}{\partial \Im(M_{i})}\Delta \Im(M_{i})\right] \, .
\end{equation}
That is, each disturbance contributes error to the differential image intensity through the change in its variance (in going from the target to the reference star) and the change in the real and imaginary parts of its mean contributing field. We can think of this as the temporal change of the total intensity inside a given ROI. 
}

\textcolor{revA}{
The error budget is based on the ensemble variance of $\Delta I$. For uncorrelated contributions $x_i$, the variance of the sum equals the sum of the variances:
\begin{equation}
\label{eq:sigma_sum}
    \sigma^2(\sum_{i} x_{i})=\sum_{i} \sigma_{i}^2 \, .
\end{equation}
It can be shown that, using this property of uncorrelated contributions, and Eq.~\ref{eq:Dif_Sum_I}, the ensemble variance of intensity changes is given by:
\begin{equation}
    \label{eq:sigma_dI_squared}
    \sigma_{\Delta I}^{2}=\sum_{i=0}^{n} \sigma_{\Delta V_{i}}^2 + (2\Re(M_{tot}))^2\sum_{i=0}^{n} \sigma_{\Re(\Delta M_{i})}^{2} + (2\Im(M_{tot}))^2\sum_{i=0}^{n} \sigma_{\Im(\Delta M_{i})}^{2} \, .
\end{equation}
}

\textcolor{revA}{
Each disturbance mechanism $i$ makes its own contribution to the total $E(x,y,t)$, say via ${\rm WFE}_{i}(x',y',t)$, where $(x',y')$ are pupil plane coordinates, and the phase of the resultant field contribution $E_{i}(x,y,t)$ will in general change by a different amount in each instance of the ensemble, with the change ranging anywhere from 0 to 2$\pi$. If this is so, then we can expect, for any mechanism $i$, that the real and imaginary parts of $\Delta M_{i}$ have similar variances over the ensemble of ROI's:
\begin{equation}
    \label{R&I_Equality}
    \sigma_{\Re(\Delta M_{i})}^2=\sigma_{\Im(\Delta M_{i})}^2=\frac{1}{2}\sigma_{\Delta M_{i}}^2
\end{equation}
}

\textcolor{revA}{
Incorporating the above observation into Eq.~\ref{eq:sigma_dI_squared} leads to the key result:
}
\begin{equation}
    \label{eq:Contrast_Stability}
    \boxed{\sigma_{\Delta I}^2=\sum_{i=0}^{n}\sigma_{\Delta V_{i}}^2+2 \lvert M_{tot}\rvert^2\sigma_{\Delta M_{i}}^2}
\end{equation}

That is, in considering an ensemble of direct imaging instances using RDI, and considering any particular region of interest, the ensemble-wide variance of the differential intensity is given by two groups of contributions: the variance changes between reference and target looks, and the mean changes. The latter are furthermore \textit{amplified} coherently by the total mean field. The three quantities that appear inside the sum are:
\begin{itemize}
    \item $\sigma_{\Delta V_{i}}^2$: the ensemble variance of \textcolor{revA}{ the change in the temporal variance      when going from target to reference}
    \item $M_{tot}$: the ensemble mean of the total field 
    \item $\sigma_{\Delta M_{i}}^2$: the ensemble variance of the temporal \textit{mean} changes in going from one observation to the other
\end{itemize}

\textcolor{revA}{
\subsection{Connecting to Contrast and Flux Ratio Noise}
}
\textcolor{revA}{
Equation~\ref{eq:Contrast_Stability} provides an estimate of the ensemble variance of the residual speckle intensity. However, contrast and flux ratio are \textit{normalized} intensities. 
Contrast is defined by Eq.~\ref{eq:contrast} as a ratio of two throughputs, but it can equivalently be calculated using the corresponding focal plane intensities. Furthermore, in practice it is easier to start with normalized intensity NI (Eq.~\ref{eq:NI}) and convert to contrast later. In Eq.~\ref{eq:NI}, the denominator is $\tau_\phi(0,0)$, which can be written as: 
\begin{equation}
    \label{eq:intensitythoughputnofpm}
    \tau_\phi(0,0)=\frac{I_\phi(0,0)}{I_{in}} \, .
\end{equation}
$I_\phi$ is the intensity in the focal plane with the focal plane mask removed when the star intensity entering the telescope is $I_{in}$.
We can now cast Eq.~\ref{eq:Contrast_Stability} in terms of $\sigma_{\Delta NI}^2$, the differential image normalized intensity ensemble variance:
\begin{equation}
    \label{eq:speckleNoise_normalizedIntensity}
    \sigma_{\Delta NI}^2=\frac{\sigma_{\Delta I}^2}{I_{\phi}^2} =  \sum_{i=0}^{n}  \left[  \sigma^2_e\biggl\{\frac{\Delta V_i}{I_{\phi}} \biggr\}  +   
    2 \cdot \frac{\langle \lvert M_{tot} \rvert^2 \rangle_e}{I_{\phi}} \cdot \sigma^2_e\biggl\{\frac{  \lvert\Delta M_i \rvert }{\sqrt{I_{\phi}}} \biggr\} \right] \, .
\end{equation}
On the right hand side, we have made more explicit the ensemble-wide mean and variance using the notation  $\langle\ldots \rangle_e$ and  $\sigma^2_e\{\ldots\}$, respectively. 
}

\textcolor{revA}{
To convert from normalized intensity to contrast, we use the NI-to-contrast conversion factor $c(u,v)$ given by Eq.~\ref{eq:c_over_NI}. It is convenient to average $c(x,y)$ azimuthally for each radial slice and thus turn it into a radial correction $\bar{c}(r)$. 
}

\textcolor{revA}{
Finally to convert the contrast variance to flux ratio noise, we have to use the conversion factor $\kappa_c$ , defined as the ratio of the flux ratio noise caused by speckle instability to the , derived elsewhere\cite{Nemati2020b} and computed using a diffraction model of  the coronagraph\cite{Krist2015b}: 
\begin{equation}
    \label{eq:kappaC}
    \kappa_c \equiv \frac{\delta \xi}{\sigma_{\Delta C}} = \frac{\tau_{pk}\ n_{core}}{\tau_{core}} \, ,
\end{equation}
where $n_{core}$ is the number of diffraction modeling pixels it takes to cover the PSF core, $\tau_{core}$ is the core throughput given by Eq.~\ref{eq:tau_core}, and $\tau_{pk}$ is the peak throughput as used in Eq.~\ref{eq:contrast}. $\tau_{pk}$ is proportional to the integrated intensity into a single modeling pixel. Note that as the modeling pixel sampling is changed, the product $\tau_{pk}\ n_{core}$ is unchanged (to first order) and neither is $\kappa_c$. 
}
 
Putting it all together, the flux ratio noise contribution from speckle noise is given by:
\begin{equation}
    \label{eq:delta_xi_cbar}
    \delta\xi = \kappa_c\cdot\bar{c}(r)\cdot\sigma_{\Delta NI} \, ,
\end{equation}
\textcolor{revA}{
where $\bar{c}(r)$ is the azimuthally averaged $c(u,v)$ from Eq.~\ref{eq:c_over_NI}.
}

To summarize, we started with fields and their variations, and derived the contrast stability statistics, and finally tied them to flux ratio noise through normalization:
\begin{equation}
    E(u,v,t) \rightarrow \sigma_{\Delta I} \rightarrow \sigma_{\Delta NI} \rightarrow \sigma_{\Delta C} \rightarrow \delta\xi
\end{equation}
 
Contrast instability $\sigma_{\Delta C}$ is calculated using statistics from the disturbances, obtained from integrated modeling runs following specific observing scenarios, and sensitivities computed using diffraction modeling of the optical system. The details on this procedure will be subject of a forthcoming paper.

\section{Calibration Errors}
\label{sect:Calibration} 

\begin{figure}[ht]
	\centering
	\includegraphics[width=0.3\columnwidth]{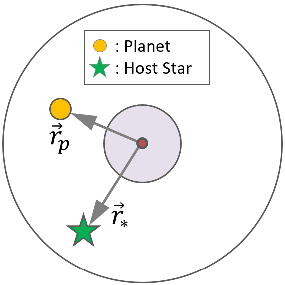}
	\caption{Flux ratio determination requires knowledge of the star flux in band. In a coronagraph, this means the star also needs to be measured off-axis, and ideally at the planet location. }
	\label{fig:CalOverview}
\end{figure}

Calibration is used to correct the final signal counts and to convert them, via Eq.~\ref{eq:xi_p}, into a flux ratio measurement. Fig.~\ref{fig:CalOverview} shows the essential aspects of flux ratio calibration. Once the planet is detected, ideally the star is brought to the same part of the field of view and its flux measured with all other settings kept the same. Under this ideal scneario, a precise way of mathematically expressing the measured flux ratio would be as the ratio of two integrals over the waveband of interest:
\begin{equation}
	\xi = \frac{\int F_p \tau_p \tau_f d\lambda}{\int F_s \tau_p \tau_f d\lambda}  	\, .
	\label{eq:FRprecise}
\end{equation}
$F_p$ and $F_s$ are the actual spectral fluxes from the planet and the host star, respectively, during the visit. The wavelength-dependent function $\tau_p$ is the conversion factor from photons entering the telescope to electrons in the region of interest (~ PSF core). The filter passband function is $\tau_f$. 

In practice, however, the planet signal processing is not completed until long after the visit, so the location of the planet to which the star could be offset pointed is not known during the visit. Therefore, the star is instead offset-pointed to a specific, well calibrated part of the field of view (within the dark hole) and its flux is measured there. For most targets, this also entails the use of a neutral density (ND) filter to avoid saturation of the detector. 

Operationally, therefore, we have the planet at some location $\vec{r}_p$ and the star at some other location  $\vec{r}_s$ when offset-pointed. We obtain a signal $S_p$ from the planet and $S_s$ from the star. The flux ratio calculation starting from these two signals is then:
\begin{equation}
	\xi = \frac{S_p(\vec{r}_p)}{S_s(\vec{r}_s)}\cdot \frac{t_s}{t_p} \cdot \frac{\int F_s \tau_s \tau_f d\lambda}{\int F_s \tau_p \tau_f d\lambda}  	\, ,
	\label{eq:FRspelledout}
\end{equation}
where we have highlighted the fact that $S_p$ and $S_s$ are obtained at different locations in the dark hole. On the right hand side, the second factor is a correction for the different integration times on the star and the planet, and the third factor is the correction for measuring the star at a different location than the planet, with a different throughput: $\tau_p$ at $\vec{r}_p$ and $\tau_s$ at $\vec{r}_s$.

The experimental version of the calibration factor $\kappa$, first introduced in Eq.~\ref{eq:kappa}, is:
\begin{equation}
	\kappa_{exp} = \frac{1}{S_s(\vec{r}_s)}\cdot \frac{t_s}{t_p} \cdot \frac{\int F_s \tau_s \tau_f d\lambda}{\int F_s \tau_p \tau_f d\lambda}  	\, ,
	\label{eq:kappaexp}
\end{equation}
With these definitions, we arrive at a working, experimental form of Eq.~\ref{eq:xi_p}:
\begin{equation}
	\xi_{exp} = S_p(\vec{r}_p) \cdot \kappa_{exp} 	\, .
	\label{eq:kappa_experimental}
\end{equation}
That is, the experimental way to obtain the flux ratio involves measuring not just the signal, but also the calibration factor $\kappa_{exp}$. Note that $S$ in  Eq.~\ref{eq:xi_p} is more explicitly called $S_p(\vec{r}_p)$ here.

The bulk of the calibration activity is to determine the third factor in Eq.~\ref{eq:FRspelledout}. This is a ratio of two integrals, and is not strictly separable into factors. To the extent, however, that we can assume they are mostly flat across the passband of interest, the integral becomes separable into factors. We therefore proceed by making a \textit{factorization approximation}. With this approximation adopted, we can rewrite the expression for the experimentally determined calibration factor as: 
\begin{equation}
	\kappa_{exp} \simeq 
	\left( \frac{\tau_{ND}(\vec{r}_s)}{S_s(\vec{r}_s)}\cdot\frac{t_s}{t_p} \right) \cdot
	\left( \frac{\tau_{CTI}(\vec{r}_s)}{\tau_{CTI}(\vec{r}_p)} \right) \cdot
	\left( \frac{\tau_{core}(\vec{r}_s)}{\tau_{core}(\vec{r}_p)} \right) \cdot
	\left( \frac{\tau_{FF}(\vec{r}_s)}{\tau_{FF}(\vec{r}_p)} \right) \cdot
	\left( \frac{\tau_{IM}(\vec{r}_s)}{\tau_{IM}(\vec{r}_p)}  \right)
 	\, .
	\label{eq:factorisation}
\end{equation}
We now label each factor above separately:
\begin{itemize}
	\item $\left( \frac{\tau_{ND}(\vec{r}_s)}{S_s(\vec{r}_s)}\cdot\frac{t_s}{t_p} \right)$: Star flux calibration, conducted at the star position $\vec{r}_s$;
	\item $\left( \frac{\tau_{CTI}(\vec{r}_s)}{\tau_{CTI}(\vec{r}_p)} \right)$: Charge 
	transfer inefficiency (CTI) calibration, relative values  $\vec{r}_s$ to  $\vec{r}_p$;
	\item $\left( \frac{\tau_{core}(\vec{r}_s)}{\tau_{core}(\vec{r}_p)} \right)$: Core throughput calibration, relative values  $\vec{r}_s$ to  $\vec{r}_p$;
	\item $\left( \frac{\tau_{FF}(\vec{r}_s)}{\tau_{FF}(\vec{r}_p)} \right) $: Flat-field calibration, relative values  $\vec{r}_s$ to  $\vec{r}_p$;
	\item $\left( \frac{\tau_{IM}(\vec{r}_s)}{\tau_{IM}(\vec{r}_p)}  \right)$: Image corrections, relative values  $\vec{r}_s$ to  $\vec{r}_p$;
\end{itemize}

How the calibration is done affects the breakdown grouping and allocations.
Since $\kappa_{exp}$ is a product of mostly independent factors, the budget for calibration can be most conveniently based on \textit{fractional} errors that add in quadrature. 
The result is a fractional error in $\kappa_{exp}$, which, by Eq.~\ref{eq:kappa_experimental}, causes the same fractional error in $\xi_{exp}$.

A detailed description of the calibrations for Roman CGI are beyond the scope of this paper, but here we provide the calibration budget and the aspects covered by each calibration.
In Table~\ref{tab:Calibrations} are listed the allocations for each of the calibrations required for the detemination of the flux ratio $\xi_{exp}$.  The last entry,
\textcolor{revA}{\textit{Bias pattern noise}}, is not a factor for $\kappa$ but is a correction that is captured in the photometry branch of the budget. 
\begin{table}[ht]
	\fontsize{9}{11}\selectfont
	\caption{Calibrations and what they capture.} 
	\label{tab:Calibrations}
	\begin{center}       
		\begin{tabular}{l c l l} 
			Calibration & Allocation & Main Focus of the calibration & Approach\\ \hline\hline
			Star flux   & 2.0\% & Stellar flux through the filter  & Star off axis at multiple locations\\
			CTI         & 2.7\% & Charge transfer inefficiency (traps) & Trap pumping \& removal algorithms\\
			Core throughput   & 2.2\% & Core throughput vs. radius & Offset pointing and raster over the DH \\
			Flat field  & 2.0\% & transmissions and reflectivities, QE & Neptune raster scanned across FOV \\
			Image correction  & 1.6\% & Detector gains, nonlinearity  &  Photon transfer curve \\
			Bias pattern noise & 1.5\% & Dark current, CIC, and fixed pattern & Periodic sets of dark frames \\
			\hline 
		\end{tabular}
	\end{center}
\end{table} 

\textit{Star Photometry}: Measures the in-band flux of the star, under the same configuration as the planet signal, but at the star location, and corrected for the ND filter that needed to be used, and the different integration time. It includes the diffractive effects of the coronagraphic masks. With the masks in place, and additionally a neutral density (ND$\sim$4) filter in place, and guiding using the observatory, the star is moved from its nominal position (center of line of site), and placed at a range of separations from the inner working angle to the outer working angle. 

\textit{Charge Transfer Inefficiency (CTI)}: There are two sources of CTI: clocking inefficiency and charge traps. Over time, the trap effect dominates on on-orbit calibrations will be necessary. To locate and identify the charge traps, the detector will be operated in trap pumping mode.\cite{Hall2014,Hall2014a}. Separately an algorithm that reconstructs the original image using the knowledge of the existing trap statistics is used to remove their effects.\cite{Massey2014}. In the error budget, this captures the error in the correction for CTI at for the (offset-pointed) star relative to the planet. 

\textit{Core Throughput}: Corrects for the difference in the core throughput for the offset-pointed star relative to the planet. This calls for a relative calibration of core throughput within the dark hole. The calibration is done by offset pointing a standard star of V-magnitude $\ge$ 10.9 using the fast streering mirror (FSM). The magnitude limit is set by the brightest object that can be offset pointed and not saturate the detector at the shortest practical exposure times. A large number of points are measured and the results are fitted to a smooth function. 

\textit{Flat Field}: Corrects for the non-diffractive part of the conversion efficiency at the offset-pointed star location relative to the planet. With the coronagraphic masks removed, an extended object, like Neptune, Uranus, or M31 is rastered over the field of view, using the fast steering mirror. The observatory pointing is maintained at lower precision without CGI control feedback. This calibration captures the optical throughput including the reflections, transmissions, as well as the detector pixel response non uniformity. 

\textit{Image Correction}: Image correction is needed for raw frames and includes the detector ``K'' gain (conversion from analog-to-digital ADU units to electrons), nonlinearity of the detector, photon-counting photometric correction, and the removal of cosmic ray hits. Photon-counting photometric corrections account for thresholding and coincidence losses, and require calibration of the EM gain. These were mentioned in Section~\ref{photoncounting} and are described in detail in (Nemati 2020).\cite{Nemati2020a}
 
\textit{Bias pattern noise}: Raw frames need to be corrected for dark current, clock induced charge, and fixed pattern noise. All three of these effects are in general spatially variable over the frame, so the correction requires the generation of a ``map'' which is subtracted from the raw frame. A fourth effect included here for photon counted frames (planet signal frames) is read noise leakage, mentioned in Section~\ref{photoncounting}. Since the map is subtracted, this calibration is not included with the rest of the items in this list in the Calibration branch, but in the detector noise box of the photometric branch. 
 
\section{The CGI Error Budget}

\begin{figure}[ht]
	\begin{center}
		\includegraphics[width=0.99\columnwidth]{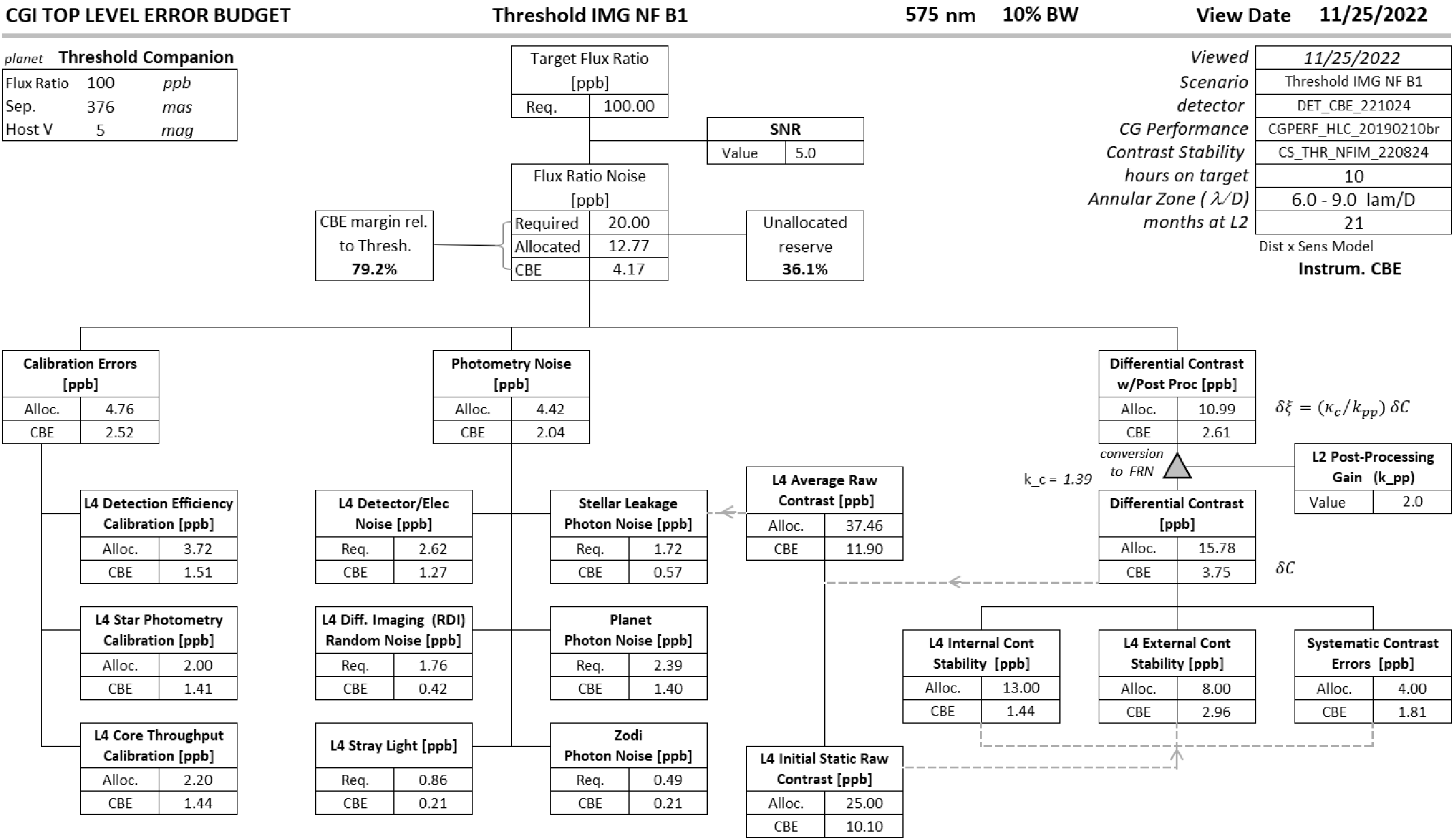}
	\end{center}
	\caption 
	{ \label{fig:EBtoplevel}
		The Roman Coronagraph Error Budget tracks flux ratio noise, and has three main branches. Each box contains the allocation (REQ) and current best estimate (CBE).} 
\end{figure}

The CGI error budget, shown in Fig.~\ref{fig:EBtoplevel}, provides the allocations (labeled either as ``Alloc.'' or ``Req.'') and ``current best estimates'' (CBE's) towards achieving a given SNR on a ``fiducial" (and hypothetical) planet in a particular observing mode. At any given reporting time, the CBE may be the result of testing or modeling or analysis. In ordinary usage, the ``best'' estimate from a model would be the nominal, the central, or the most probable value. However, it has become common engineering practice in the field in some circumstances to apply a multiplicative ``model uncertainty factor'' (MUF) to some disturbances and sensitivities.
The sensitivities in this context are the sensitivities of errors to disturbances. Following this practice, CGI uses some MUF's both on disturbances and sensitivities that go into the FRN CBE's. The CGI MUF policy is described in detail elsewhere.\cite{Krist2023End-to-endCoronagraph}

 MUF's are always greater than 1, so their use constitutes a measure of conservatism in assessing performance. Thus, the meaning of ``best'' in the context of CBE's should be taken as ``most appropriate'' to the current understanding of errors. 

In the threshold direct imaging budget, the fiducial planet is one with a flux ratio of $1\cdot10^{-7}$ orbiting a star with apparent magnitude V=5, with a separation of 376 mas, being observed in the mode NFOV Band 1 IMG (see Table~\ref{tab:CGIobsModes}).  
This choice places the fiducial planet at a working angle of $7.5~ \lambda/D$ in Band 1. It is assumed that RDI operations have been done with 10 hours on the target star, with 70\% duty factor (i.e. 7 hrs of usable integration time). The duty factor accounts for the useful fraction of the time from the standpoint of reaction wheel jitter. Post processing of the differential image is assumed to have been completed and the planet signal is assumed to be extracted via aperture photometry. 

In this threshold error budget, the required SNR is 5, to be reached within an allocated integration time of 10 hrs on the target star. The flux ratio and SNR requirements imply a maximum allowed flux ratio noise of $2\cdot10^{-8}$ (i.e. 20 parts per billion, or 20~ppb). 

As is conventional practice, the error budget allocations are made top-down while the CBE's are bottom-up. However, in our implementation, we have structured the \textit{calculations} in the error budget such that they are bottom-up for both the allocations and current best estimates. For CBE's this is natural, since subsystems report their CBE's in their native units, and those numbers need to be converted to FRN units. We have chosen to use bottom-up for requirements/allocations (REQ) also, since the conversions and methodology is already in place for CBE's, and this allows setting requirements at the native units level and only then flowing them up to the error budget. 

We track both the unallocated reserve and the margin. Unallocated reserve measures how much of the total available FRN (20 ppb) is left unallocated. If additional error sources are discovered in one area, the reserve provides a ``cushion" against disturbing already-established requirements in other areas. Margin is similarly defined, but applies to the CBE's: it measures what fraction of total allowable error remains after accounting for all the currently known estimates. The formula for both margin and reserve can be written as:
\begin{equation}
	R = 1-\frac{x}{a}  	\, ,
	\label{eq:resmar}
\end{equation}
where $R$ stands for reserve in the case of the allocations (REQ's) roll-up, and it stands for margin in the case of current best estimates (CBE's) roll up. In each case, $x$ is the roll-up and in both cases $a$ is the total available allocation (20 ppb for the Threshold error budget).

In Section~\ref{sec:Signal} we gave the relation between the signal and the flux ratio in Eq.~\ref{eq:xi_p}, which involved the flux ratio factor $\kappa$ (Eq.~\ref{eq:kappa}).
The error in $\xi$ comes from error either in $\kappa$ or $S$:
\begin{equation}
\label{eq:FR_error}
 \delta\xi  = \delta\kappa\ S \oplus \kappa\ \delta S  \, ,
 \end{equation}
where the $\oplus$ symbol refers to the quadrature sum (or root sum square) operation, applicable to independent errors.  The first term on the right hand side concerns calibration errors when converting the final extracted signal counts from the processed differential image to flux ratio. The second term, which is dominant, includes all the effects that cause error in the measurement of the extracted signal counts. In the CGI error budget, this is separated into two major categories: photometric noise and noise due to contrast instability.  
Thus, at the top level the error budget has three branches.

As noted earlier, flux ratio noise (FRN) is the metric for the Roman CGI error budget, with each term $\sigma_i$ related to a corresponding FRN term $\xi_i$ through Eq.~\ref{eq:FRNcorrespondence}. 
So, allocations are made in FRN units (ppb) and the current best estimates are computed first in terms of standard deviation of counts, and then converted to FRN via  Eq.~\ref{eq:FRNcorrespondence}. 

We now will give an overview of the contents of the error budget, branch by branch, from left to right.  

\subsection{Calibration Errors Branch}
The individual calibration categories where described in Section~\ref{sect:Calibration}. The mapping of those items is as follows. 
\begin{itemize}
\item \textit{Detection Efficiency Calibration}: this includes Flat field, Charge transfer inefficiency, and image corrections. 
\item \textit{Star Photometry Calibration}: only star photometry.  
\item \textit{Core Throughput Calibration}: only core throughput.
\end{itemize}
In all the above cases, the fractional errors in the calibration directly translate to fractional errors in FRN. So a 1\% fractional error in a calibration of a 100 ppb target is a 1 ppb FRN contribution.

\subsection{Photometry Noise Branch}
These errors are computed by taking the lower level errors, expressed in counts (electrons) and converting then to FRN using the calibration factor $\kappa$ and Eq.~\ref{eq:FRNcorrespondence}.
The boxes here are:
\begin{itemize}
	\item \textit{Detector/Electronic Noise}: this includes two types of contributions:
	\subitem Detector Noise: this includes dark current, CIC, per Eq.~\ref{eq:PCdetectorNoise}.
	\subitem Bias Pattern Noise: this is the last item in the table in Section~\ref{sect:Calibration}.
	\item \textit{Differential Imaging Random Noise}: these are the contributions covered in Section~\ref{sect:RDInoise}.
	\item \textit{Stellar Leakage Photon Noise}: this is the speckle shot noise, given by Eq.~\ref{eq:speckleShot}. The contrast $C$ that goes into this is computed and corrected for the C/NI difference via Eq.~\ref{eq:c_over_NI}.
	\item \textit{Planet Photon Noise}: planet shot/Poisson noise given by Eq.~\ref{eq:planetShot}.
	\item \textit{Zodi Photon Noise}: combined local and exo zodi shot noise given by Eq.~\ref{eq:localZodiRate} and Eq.~\ref{eq:exoZodiRate} converted according to Eq.~\ref{eq:localZodiShot} .
	\item \textit{Stray light}: These contribute by shot noise and are computed via external models. Sources of stray light are:
	\subitem Luminescence (primarily Fluorescence) of transmissive optics
	\subitem Scattered light from various optics, starting from the telescope to the coronagraph. 	
\end{itemize}
\subsection{Differential Flux Ratio with Post Processing Branch}
This branch contains the effect of average contrast and its instability, as sketched in Section~\ref{sect:Cstabilty}. At the top of the branch, there are two conversions that need explanation. The top box, called Differential Contrast with Post Processing includes two factors that are applied to the box below it:
\begin{itemize}
	\item \textit{Post Processing Gain}: this is a noise reduction factor that is expected from additional post processing beyond what is assumed in the error budget. This post processing would be offered by scientists working with the CGI data. However, as alluded to in Section~\ref{sect:conops}, ADI may confer a reduction factor of  $\sim\sqrt{2}$, leaving an additional $\sim\sqrt{2}$ from other post processing to realize this assumed gain. 
	\item \textit{$\kappa_c$}: This is the conversion factor from contrast to FRN, given by Eq.~\ref{eq:kappaC}. Regarding this conversion factor $\kappa_c$, it bears noting that, while its value is numerically close to unity and it is dimensionless, it converts an \textit{instrument} parameter (contrast or its noise) into an \textit{observational} metric (flux ratio or its noise). In other words, below this point we go from flux ratio to contrast, even though the units remain the same (ppb).  
\end{itemize}

The remaining boxes are:
\begin{itemize}
	\item \textit{Differential Contrast}: These are computed using the methodology of Section~\ref{sect:Cstabilty}. Below this box there are three that divide the errors according to whether the errors are from noise sources internal to CGI, external to CGI (the rest of Roman) or are systematics such as wavefront sensor cross talk. 
	\item \textit{Initial Static Raw Contrast}: this is the contrast before disturbances and drifts, as best obtained from digging the dark hole. It corresponds to the initial field $E_0$ and is coherent with the disturbance fields. 
	\item \textit{Average Raw Contrast}: This is the total contrast that includes coherent and incoherent contributions. The total contrast that goes into \textit{stellar leakage photon noise} is based on this contrast. 
\end{itemize}

\section{Validation of the Analytical Model}

The analytical model described here has been tested against the CGI integrated model. Here we describe the CGI integrated model, the test criteria, and the results of the validation test.

\subsection{The CGI Integrated Model}
\label{Sec:Validation}

Simulations of the two-dimensional Roman coronagraphic field over time are generated using a combination of finite element and diffractive optical modeling.\cite{Krist2015b,Krist2018}

STOP (structural, thermal, optical performance) models are used to predict the changes in the wavefront and optical alignment due to thermal response to solar incidence variations during the defined observing scenario. 

The thermal response is computed using Thermal Desktop and the resulting structural distortions derived with NASTRAN. The locations of the optical surfaces are then input into Code V to predict via ray tracing the wavefront changes and relative displacements of critical surfaces (e.g., pupils). Until recently, the STOP models ended at the entrance to CGI (the pupil image at the fast steering mirror), so changes internal to the instrument were not captured; the models now include all elements of CGI, including the thermal response to electronics. The system drifts are reported at time steps of a few minutes.

\begin{figure}[ht]
    \centering
    \includegraphics[width=0.95\columnwidth]{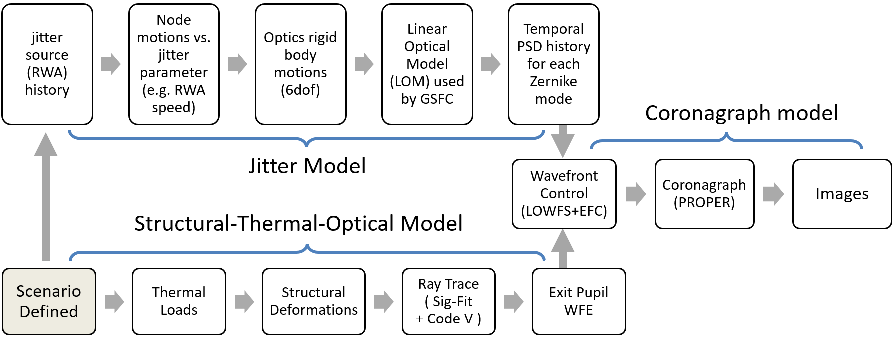}
    \caption{The CGI Integrated Model.}
    \label{fig:IntegModel}
\end{figure}

Parallel to the STOP model is a dynamical model that predicts the pointing errors caused by vibrations induced by the reaction wheels. The wheel speeds versus time are tailored for the observing scenario, including slews and rolls. Constraints are applied to ensure that the speeds are not so high as to induce large vibrations beyond the ability of the low order wavefront sensor to sense and control. The vibrations in the optics are represented as power spectral density (PSD) descriptions of wavefront error versus temporal frequency for low-order Zernike aberrations at sub-second time steps through the scenario. The tip and tilt terms, which represent pointing errors, are filtered by a function that represents the LOWFS response. The net result is a set of low-spatial-order, high-temporal-frequency (jitter) aberrations versus time step, including residual pointing errors.

The slow wavefront and alignment drifts from the STOP model and rapid wavefront and pointing jitters from the dynamics model are inputs into the optical diffraction model that computes the coronagraphic images. The telescope and CGI system are represented as an unfolded series of optical elements, each with realistic but synthetic fabrication errors appropriate for the type of optic. In later versions of the error budget, measured optical errors will be included as they become available. As the field propagates through the system, these errors will alter both its amplitude and phase, depending on where the optic is (most significantly, close or far from a pupil). This requires diffraction algorithms that propagate the beam from optic to optic. The open source PROPER propagation software library is used for this.\cite{Krist2007PROPER:IDL,PROPERHttp://proper-library.sourceforge.net}
It includes routines for Fresnel propagation and representation of complex masks and deformable mirrors.

The PROPER model is used with wavefront sensing and control (WFS/C) algorithms to derive DM patterns that create the dark hole around the star. Prior to control, the aberrated starlight that is not suppressed by the coronagraphic masks creates a background contrast of about $10^{-4}$. After applying the iterative WFS/C, the dark hole contrast improves down to $\sim10^{-9}$, so that instrumentally-generated speckles are of similar brightness to the anticipated extra-solar giant planets. This establishes the initial conditions at the beginning of the observing scenario.

At each time step, the STOP-predicted wavefront changes are added to the PROPER model and propagated to the focal plane mask, where the reflected signal is sent to the LOWFS model. This measures the low-order aberration change, including sensing noise, and then determines the adjustment to the DM necessary to compensate (with DM gain noise and finite stroke resolution included). This correction is added to the following time step. The LOWFS-corrected wavefront is then propagated through the rest of the system and onto the detector. The fields computed at STOP time steps are interpolated to finer jitter-timescale steps.

The rapid pointing and wavefront jitter effects are included at the final image by incoherently adding appropriate pre-computed electric field changes to the image plane field with weighting based on the predicted jitter amount. The final image is then sent to a detector model that adds noise and artifacts (e.g., cosmic rays).

\subsection{Validation Observing Scenario and Comparison Metric}

The measurement noise part of the analytical model follows well-known procedures and was  tested previously, so the emphasis in the comparison was on the contrast instability part of the model.
We chose the direct imaging case for the comparison, and made the simplifying assumptions that there were 1) no cosmic ray hits, 2) no hot pixels, and 3) no charge traps. 
In the observing scenario chosen (CGI observing scenario number OS6), there was a relatively bright reference star (V=1.9) and a median-brightness CGI target star (V=5). 

\begin{figure}[ht]
    \centering
    \includegraphics[width=0.7\columnwidth]{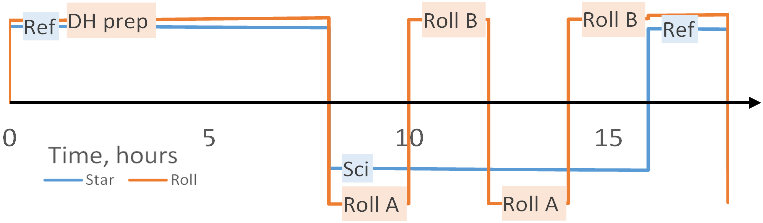}
    \caption{CGI Observing Scenario 6 (OS6) chosen for the comparison.}
    \label{fig:OS6}
\end{figure}

OS6 involves both RDI chops and roll chops. Of every 10 hours, 2 hours are spent on the bright reference star, and 8 hours on the target star. Every 2 hours a roll maneuver of $\pm$26 degrees is done, placing the roll state at $\pm 13$ degrees relative to the zero solar angle state (Fig.~\ref{fig:RDI_ADI}). The OS6 timing is shown in in Fig.~\ref{fig:OS6}. 
In the model,
\textcolor{revA}{ these maneuvers are reflected in changes in  insolation (solar radiative heating) and reaction wheel speeds with their vibration spectra.   }
The analytical model was given the disturbance statistics, as defined in the previous section, for the same scenario. 

The integrated model result is a summed image from the reference observation, and one from the target observation. Since the objective is to compare the noise, no signal (planet) is included in the target frames. The RDI image is the difference of these images, after correcting the reference image for the brightness ratio of the target relative to the reference. 

\begin{figure}[ht]
    \centering
    \includegraphics[width=0.9\columnwidth]{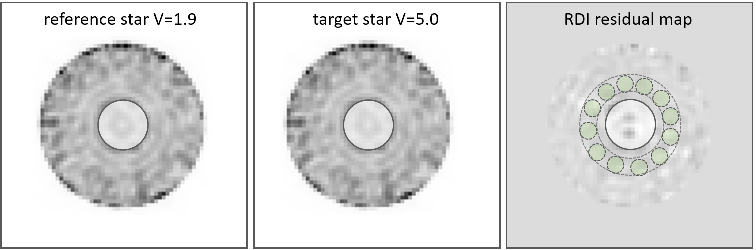}
    \caption{Reference, target, and residual images in integrated model output. The evaluation annular zone is shown on the residual map.}
    \label{fig:IMRDIresid}
\end{figure}

On the integrated model side, for the signal extraction method, a simple aperture photometry scheme was assumed, and the rms counts obtained from the RDI image, over the designated ensemble of photometric apertures with similar statistics (i.e. an annular zone) was taken as the estimated noise (Fig.~\ref{fig:IMRDIresid}). For both the integrated model and analytical model cases, the noise was converted to FRN using the calibration factor, $\kappa$ (Eq.~\ref{eq:kappa}).

\subsection{Disturbances}
For the comparison, the disturbances included were 
\begin{itemize}
    \item thermally induced wavefront error changes upstream of the coronagraph 
    \item thermally induced shear of the beam upstream of the coronagraph
    \item reaction wheel actuator (RWA) induced vibration
    \item attitude control system (ACS) tracking errors
    \item thermally induced wavefront errors and shear internal to the coronagraph
    \item PMN deformable mirror response (ferroelectricity) dependence on temperature
    \item PMN deformable mirror `creep' (asymptotic approach to commanded shape).
\end{itemize}
Some of these `primary' disturbances are corrected by the coronagraph, and it is the post-correction error (`secondary' disturbances) that matters in those cases. The secondary disturbances include response to dynamic terms,      tip-tilt correction, focus correction, and DM Z5-11 correction, and sensing noise, electronic noise  and quantization noise.

\subsubsection{Processing for the Comparison}

Using the OS6 speckle time series, we created detector frames, including the appropriate stellar flux, throughput losses, and detector noise. 
These frames were then photon counted via thresholding, as described in 
Section~\ref{photoncounting}, and corrected to first order using the approach in (Nemati 2020).\cite{Nemati2020a} 
For each RDI (target-reference) chop, a differential image was formed, after normalizing for flux difference per Eq.~\ref{eq:RDI_difference}. These differential images were co-added for the (two) chops to make a total differential image. 

A radial slice was then defined, near the IWA of the dark hole, 
\textcolor{revA}{
and a number of photometric apertures were selected around the azimuth. 
These photometric apertures were chosen as 5 contiguous pixels since $m_{pix}$ is about 5 for the HLC Band1 imaging mode, as noted in Section~\ref{detectorNoiseSection}. 
These photometric apertures constitute the regions of interest (ROI's) for would-be planets and they are core-sized.  }

\begin{figure}[ht]
    \centering
    \includegraphics[width=0.9\columnwidth]{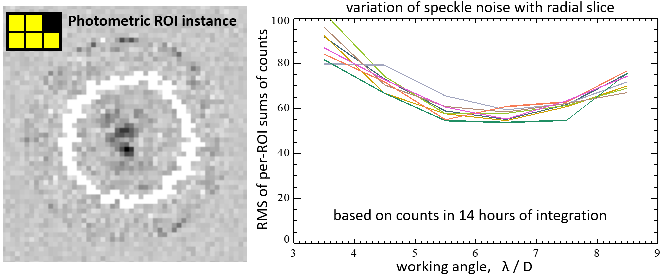}
    \caption{The ROI selections around the azimuth for one ROI shape and orientation is shown on left, inset. On right we plot, for the eight possible ROI shapes, the rms over all ROI's of the summed counts within the ROI. A single instance of ROI shape is shown on the left. The mean of these curves is used for the comparison with the analytical model.}
    \label{fig:EBIMroi}
\end{figure}

The error, in counts (detected photo-electrons), was calculated as the rms, over the ensemble of ROI's, of the sum of the counts within each ROI. Fig.~\ref{fig:EBIMroi} shows the result. Given that the core is only $\sim5$ pixels in size, the ROI is not circularly symmetric. The left map of ROI's shows in an inset the orientation of the ROI. The right plot shows a graph for each unique orientation of the ROI, and the average of these graphs was used as the best estimate of the error.
The conversion to FRN was done, per Eq.~\ref{eq:FRNcorrespondence}, by multiplying by the calibration factor $\kappa$ (Eq.~\ref{eq:kappa}).

\begin{figure}[ht]
    \centering
    \includegraphics[width=0.7\columnwidth]{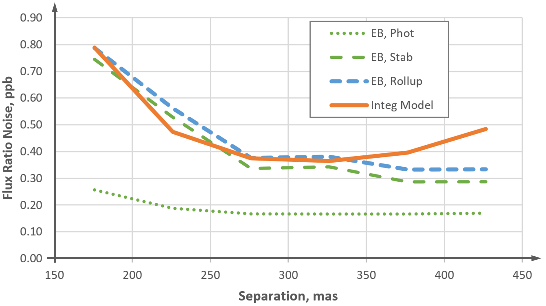}
    \caption{\textcolor{revA}{Comparison of the computed error as a function of angular separation for the integrated model (solid, thick, orange) vs. the Error Budget rollup (dashed, thick, blue). The latter is the quadrature sum of the photometric noise (thin, dotted, green) and the contrast stability (thin, dashed green). }}
    \label{fig:EBIM}
\end{figure}

For the analytical estimate, the sensitivity was computed for different annular zones, and the disturbance was taken from the OS6 specification. The contrast stability error was computed per Section~\ref{sect:Cstabilty}. The photometric error was computed using Eq.~\ref{eq:refnoise_total}. The results are shown in Table~\ref{tab:ROIebim} \textcolor{revA}{and summarized in Fig.~\ref{fig:EBIM} }for each working angle (WA) annular slice. The fractional error is computed as $(\rm{EB}-\rm{IM})/\rm{IM}$. We find that the rms error across all annuli is 16\%. For a model validation of this type this is considered quite an acceptable level of fidelity, and particularly so in light of the fact that there are many measures of conservatism already built in, such as the use of MUF's in computing CBE's. 

\begin{table}[ht]
\centering
\tiny
\caption{Comparison of the estimate of flux ratio noise (FRN) according to the analytical model (EB, for error budget) versus the integrated model (IM). In the EB case the two columns noted are contrast stability and photometry, which are two of the branches of the error budget as shown in Fig.~\ref{fig:EBtoplevel}. The third branch is not used in the comparison since it does not pertain to signal extraction errors, which was the scope of the comparison.}
\vspace{10pt}
\resizebox{\textwidth}{!}{%
\begin{tabular}[ht]{c c | c c c | c c c |c}
WA & sep  & EB C. Stab.  & EB Phot.  & EB total  & IM  &  $\kappa$  & IM     & Fractional   \\
$\lambda/D$ & mas & ppb & ppb & ppb & $\langle$ rms cts $\rangle$   & ppb/ct &  ppb & Error \\
\hline\hline
\rule[-1ex]{0pt}{3.5ex}  3-4 & 175.7 & 0.74 & 0.26 & 0.79 & 89.5 & $8.82\cdot 10^{-3} $ & 0.79 & -0.3\% \\
\rule[-1ex]{0pt}{3.5ex}  4-5 & 225.9 & 0.53 & 0.19 & 0.56 & 71.9 & $6.58\cdot 10^{-3} $ & 0.47 &   19\% \\
\rule[-1ex]{0pt}{3.5ex}  5-6 & 276.1 & 0.34 & 0.17 & 0.38 & 58.9 & $6.37\cdot 10^{-3} $ & 0.38 &  0.2\% \\
\rule[-1ex]{0pt}{3.5ex}  6-7 & 326.2 & 0.34 & 0.17 & 0.38 & 56.9 & $6.41\cdot 10^{-3} $ & 0.36 &  4.5\% \\
\rule[-1ex]{0pt}{3.5ex}  7-8 & 376.4 & 0.29 & 0.17 & 0.33 & 61.2 & $6.47\cdot 10^{-3} $ & 0.40 &  -16\% \\
\rule[-1ex]{0pt}{3.5ex}  8-9 & 426.6 & 0.29 & 0.17 & 0.33 & 72.6 & $6.67\cdot 10^{-3} $ & 0.48 &  -31\%
\end{tabular}}
\label{tab:ROIebim}
\end{table}

\section{Summary and Conclusion}

We have described in detail the analytical model that undergirds the Roman coronagraph flux ratio noise error budget. The analytical model includes both relatively conventional components, used in the photometry branch, as well as a more novel methodology to model speckle noise. We conducted a validation test of this analytical model, based on a typical CGI observing scenario, against our full propagation model which was validated against testbeds. We find that over most of the dark hole the analytical model agrees to better than 20\% with the full propagation model. This methodology has since been adopted by the Giant Magellan Telescope (GMT) project for their high contrast imaging error budget, and is well suited for future space telescopes that would aim to detect Earth-like planets in the habitable zones of nearby Sun-like stars.


\subsection* {Code, Data, and Materials Availability} 
This paper presents a methodology for analytical modeling and does not rely on any collected data. The PROPER optical model\cite{Krist2007PROPER:IDL} which was used in the validation process described in Section~\ref{Sec:Validation} can be downloaded from Sourceforge.\cite{PROPERHttp://proper-library.sourceforge.net}

\acknowledgments 
The authors would like to acknowledge Dr. Wesley A. Traub, who inspired the analytical approach presented here, and Dr. Charley Noecker of JPL for his early contributions to the CGI error budget. The work described in this paper was supported by the Roman Space Telescope Coronagraph Instrument project and  carried out for the Jet Propulsion Laboratory, California Institute of Technology, under a contract with the National Aeronautics and Space Administration.


\bibliography{references}   
\bibliographystyle{spiejour}   


\vspace{2ex}\noindent\textbf{Bijan Nemati} received his PhD in high energy physics from the University of Washington in 1990. Since 2001 he has been working on modeling and technology development of advanced space-based astronomical instruments, first at the Jet Propulsion Laboratory and recently at Tellus1 Scientific, a company he founded. On the Roman coronagraph, he has served as detector lead, modeling lead, and system engineer in charge of the coronagraph error budget.  

\vspace{2ex}\noindent\textbf{John Krist} 
 is a research scientist at Jet Propulsion Laboratory. He is the Integrated Modeling team lead for the Roman Space Telescope coronagraph. He specializes in optical diffraction modeling of space telescopes and coronagraphic imaging of circumstellar disks with the Hubble and James Webb space telescopes.

\vspace{2ex}\noindent\textbf{Ilya Poberezhskiy} 
 received his PhD in Electrical Engineering from the University of California, Los Angeles (UCLA). Since 2004, he has worked at the Jet Propulsion Laboratory on development and testing of optical instruments and subsystems for space applications. On the Roman Coronagraph Instrument project, he initially led the technology development team that demonstrated coronagraph technology readiness for flight implementation, and later served as the lead Project System Engineer.

\vspace{2ex}\noindent\textbf{Brian Kern} received his PhD in astronomy from Caltech in 2002.  Since 2005 he has been at the Jet Propulsion Laboratory, working on the High Contrast Imaging Testbed, designing, modeling, and performing high contrast coronagraph demonstrations, and since 2013 on Roman CGI as a laboratory demonstration lead and on the system engineering team.
 

\end{spacing}
\end{document}